\documentclass[twocolumn]{aastex631}

\usepackage[T1]{fontenc}
\usepackage{ae,aecompl}
\usepackage{natbib}

\usepackage{graphicx}	
\usepackage{amsmath}	
\usepackage{amssymb}	
\usepackage{bm}		
\usepackage{booktabs}
\usepackage{mathtools}
\usepackage{makecell}
\usepackage{bm}
\usepackage{float}
\usepackage[utf8x]{inputenc} 
\usepackage{color,soul}

\newcommand{\angstrom}{\textup{\AA}}
\newcommand{\fesc}{\ensuremath{f_{\rm esc}}}
\newcommand{\fout}{\ensuremath{\langle f_{900}/f_{1500}\rangle_{\rm out}}}
\newcommand{\fobs}{\ensuremath{\langle f_{900}/f_{1500}\rangle_{\rm obs}}}
\newcommand{\ziii}{\ensuremath{z\sim3}}
\newcommand{\wlya}{\ensuremath{W_{\lambda}}(\rm Ly\ensuremath{\rm \alpha})}

\newcommand{\luv}{\ensuremath{L_{\rm UV}}}

\newcommand{\fesca}{\ensuremath{f_{\rm esc,abs}}}

\newcommand{\vjh}{\ensuremath{V_{606}J_{125}H_{160}}}

\newcommand{\mstar}{\ensuremath{\rm M_{\rm *}}}

\newcommand{\vsep}{v\ensuremath{_{\rm sep}}}
\newcommand{\zlyar}{\ensuremath{z_{\rm Ly\alpha,red}}}
\newcommand{\zlyab}{\ensuremath{z_{\rm Ly\alpha,blue}}}
\newcommand{\vlyar}{v\ensuremath{_{\rm Ly\alpha,red}}}
\newcommand{\vlyab}{v\ensuremath{_{\rm Ly\alpha,blue}}}
\newcommand{\zsys}{\ensuremath{z_{\rm sys}}}
\newcommand{\zlis}{\ensuremath{z_{\rm LIS}}}
\newcommand{\vlma}{v\ensuremath{_{\rm Ly\alpha-LIS}}}
\newcommand{\vlyarh}{v\ensuremath{_{\rm Ly\alpha,red,high}}}
\newcommand{\vlyarl}{v\ensuremath{_{\rm Ly\alpha,red,low}}}
\newcommand{\vlmah}{v\ensuremath{_{\rm Ly\alpha-LIS,high}}}
\newcommand{\vlmal}{v\ensuremath{_{\rm Ly\alpha-LIS,low}}}
\newcommand{\fcen}{\ensuremath{f_{\rm cen}}}

\newcommand{\ksep}{KLCS\ensuremath{_{\rm sep}}}
\newcommand{\klya}{KLCS\ensuremath{_{\rm Ly\alpha,red}}}
\newcommand{\klma}{KLCS\ensuremath{_{\rm Ly\alpha-LIS}}}

\newcommand{\lya}{Ly$\rm \alpha$}
\newcommand{\oiii}{[O~\textsc{iii}]\ensuremath{\lambda\lambda4959,5007}}
\newcommand{\oii}{[O~\textsc{ii}]\ensuremath{\lambda\lambda3726,3729}}
\newcommand{\hb}{H$\rm \beta$}

\shorttitle{Ly$\alpha$ profile shape and \fesc{}}
\shortauthors{Pahl et al.}

\begin{document}
	
	\title[Ly$\alpha$ profile shape and \fesc{}]{Ly$\alpha$ profile shape as an escape-fraction diagnostic at high redshift}
	
	\author[0000-0003-4464-4505]{Anthony Pahl}
	\altaffiliation{Carnegie Fellow}
	\affiliation{The Observatories of the Carnegie Institution for Science, 813 Santa Barbara Street, Pasadena, CA 91101, USA}
	\affiliation{Department of Physics and Astronomy, University of California, Los Angeles, CA 90095, USA}
	
	\author{Alice Shapley}
	\affiliation{Department of Physics and Astronomy, University of California, Los Angeles, CA 90095, USA}
	
	\author{Charles C. Steidel}
	\affiliation{Cahill Center for Astronomy and Astrophysics, California Institute of Technology, MC249-17, Pasadena, CA 91125, USA}
	
	\author{Naveen A. Reddy}
	\affiliation{Department of Physics and Astronomy, University of California Riverside, Riverside, CA 92521, USA}
	
	\author{Yuguang Chen}
	\affiliation{Department of Physics and Astronomy, University of California Davis, 1 Shields Avenue, Davis, CA 95616, USA}
	
	\author{Gwen C. Rudie}
	\affiliation{The Observatories of the Carnegie Institution for Science, 813 Santa Barbara Street, Pasadena, CA 91101, USA}
	
	\begin{abstract}
		While the shape of the \lya{} profile is viewed as one of the best tracers of ionizing-photon escape fraction (\fesc{}) within low redshift ($z\sim0.3$) surveys of the Lyman continuum, this connection remains untested at high redshift. Here, we combine deep, rest-UV Keck/LRIS spectra of 80 objects from the Keck Lyman Continuum Spectroscopic Survey with rest-optical Keck/MOSFIRE spectroscopy in order to examine potential correlations between \lya{} profile shape and the escape of ionizing radiation within $z\sim3$ star-forming galaxies. 
		We measure the velocity separation between double-peaked \lya{} emission structure (\vsep{}), between red-side \lya{} emission peaks and systemic (\vlyar{}), and between red-side emission peaks and low-ionization interstellar absorption lines (\vlma{}).
		We find that the IGM-corrected ratio of ionizing to non-ionizing flux density is significantly higher in KLCS objects with lower \vlyar{}. We find no significant trend between measures of ionizing-photon escape and \vlma{}. We compare our results to measurements of $z\sim0.3$ ``Green Peas" from the literature and find that KLCS objects have larger \vsep{} at fixed \vlyar{}, larger \fesc{} at fixed \vlyar{}, and higher \vlyar{} overall than $z\sim0.3$ analogs. We conclude that the \lya{} profile shapes of our high-redshift sources are fundamentally different, and that measurements of profile shape such as \vlyar{} map on to \fesc{} in different ways. We caution against building reionization-era \fesc{} diagnostics based purely on \lya{} profiles of low-redshift dwarf galaxies. Tracing \vsep{}, \vlyar{}, and \fesc{} in a larger sample of $z\sim3$ galaxies will reveal how these variables may be connected for galaxies at the epoch of reionization. 
	\end{abstract}
	

	\section{Introduction} \label{sec:intro}
	
	Cosmic reionization represents a significant milestone in the evolution of the Universe, in which Hydrogen in the intergalactic medium (IGM) transitioned from neutral to ionized. Based on the emergent UV spectrum of distant quasars, reionization is thought to end at $z\sim5.5-6$ \citep{fanObservationalConstraintsCosmic2006,beckerMeanFreePath2021}. While there are multiple constraints on the timeline of reionization, the relative contribution of ionizing light from astronomical sources, and thus the chief drivers of this process, remain in debate. Some works have shown that star-forming galaxies provide the majority of the ionizing emissivity at $z\gtrsim6$, considering quasars have relatively small number densities during reionization \citep{inoueEscapeFractionIonizing2006,jiangFinalSDSSHighredshift2016,parsaNoEvidenceSignificant2018}, but which population of star-forming galaxies are most critical to reionization, and the detailed physics of ionizing photon production and escape, remain unsettled questions \citep[e.g.,][]{finkelsteinConditionsReionizingUniverse2019,naiduRapidReionizationOligarchs2020}.
	
	In order to constrain the evolving ionizing emissivity in the Universe, we can parameterize the quantity using three variables: the cosmic star-formation rate density (commonly measured as the UV luminosity density $\rho_{\rm UV}$), the ionizing photon production efficiency $\xi_{\rm ion}$, and the escape fraction of ionizing photons {\fesc} \citep{robertsonCosmicReionizationEarly2015}. With the advent of the \textit{James Webb Space Telescope} (\textit{JWST}), high-quality near-IR observations have been acquired for a large number of reionization-era galaxies, which has led to improved constraints on both $\rho_{\rm UV}$ \citep[e.g.,][]{finkelsteinCoevolutionAGNStarforming2022,donnanEvolutionGalaxyUV2023,harikaneComprehensiveStudyGalaxies2023} and $\xi_{\rm ion}$ \citep[e.g.,][]{tangJWSTNIRSpecSpectroscopy2023} out to $z\sim9-12$. The gradual decline in $\rho_{\rm UV}$ and the elevated $\xi_{\rm ion}$ at higher redshifts seen in early \textit{JWST} surveys appear to indicate that sufficient ionizing photons are \textit{produced} in star-forming galaxies to drive reionization, but only those that \textit{escape} their source galaxies will ultimately affect the ionization state of the IGM.
	
	Determining {\fesc} directly from photometric or spectroscopic surveys is virtually impossible for reionization-era sources, even with \textit{JWST}, as the transmission of Lyman continuum (LyC) photons through the marginally-neutral IGM drops off precipitously past $z\sim4$ \citep{vanzellaDetectionIonizingRadiation2012}. Thus, significant observational effort has been expended to understand the average ionizing signal from different populations of galaxies at lower redshifts, and uncovering indirect tracers of \fesc{} for use at $z\gtrsim6$ when reionization is in progress. These efforts include surveys at $z\sim0.3$ using the Cosmic Origins Spectrograph on the \textit{Hubble Space Telescope} (\textit{HST}/COS), where observations of LyC-leaking, compact, star-forming galaxies (``Green Peas") act as purported analogs to high-redshift sources \citep[e.g.,][]{izotovDetectionHighLyman2016,izotovLowredshiftLymanContinuum2018,izotovLymanContinuumLeakage2021}. Further success has been found by including extreme optical emission-line ratios or blue UV slopes as selection criteria, thus probing a wider range of parameter space  \citep{fluryLowredshiftLymanContinuum2022}. Closer to the epoch of reionization at $z\sim3-4$, the transmission of the LyC through single IGM sightlines is inherently uncertain, making \fesc{} constraints for \textit{individual} galaxies indeterminate. Thus, statistical samples must be utilized, as the average transmission across many sightlines can be precisely constrained. Both photometric surveys of LyC, such as those using \textit{HST}/WFC3 \citep[e.g.,][]{fletcherLYMANCONTINUUMESCAPE2019,begleyVANDELSSurveyMeasurement2022}, and those using deep, rest-UV spectroscopic measurements \citep[][]{marchiNewConstraintsAverage2017,marchiLyaLymanContinuumConnection2018,steidelKeckLymanContinuum2018,pahlUncontaminatedMeasurementEscaping2021} have presented a consensus: $\sim L^*$ star-forming galaxies have an average {\fesc} of $5-10\%$ at $z\sim3-4$.
	
	Within these surveys, the Ly$\alpha$ spectral feature appears to be a remarkably successful indicator of escaping ionizing light. Direct correlations have been found between \fesc{} and the equivalent width of Ly$\alpha$ at $z\sim3$ \citep{marchiLyaLymanContinuumConnection2018,steidelKeckLymanContinuum2018,pahlUncontaminatedMeasurementEscaping2021}. In ``down the barrel" observations, both LyC and the resonant Ly$\alpha$ photons will be attenuated by neutral-phase gas in the interstellar and circumgalactic medium (ISM and CGM, respectively), intertwining the fate of these photons. At lower redshift, the shape of the emergent Ly$\alpha$ profile has been demonstrated to strongly predict LyC leakage. When Ly$\alpha$ features double-peaked structure, the velocity separation of these peaks is inversely correlated with \fesc{}. This correlation is the tightest observed between galaxy properties (e.g., O$_{32}$, \mstar{}) and \fesc{} \citep{izotovLowredshiftLymanContinuum2018,izotovLymanContinuumLeakage2021}. In models where neutral gas is distributed in a spherical shell around a galaxy, shells with lower column densities result in smaller peak separation of Ly$\alpha$, explaining the connection to \fesc{} \citep{verhammeUsingLymanDetect2015,orlitovaPuzzlingLymanalphaLine2018}. The distribution of neutral gas around star-forming galaxies is likely more complex and clumpy, with some regions being entirely opaque and some and transparent to LyC photons \citep{rudieColumnDensityDistribution2013,reddyConnectionReddeningGas2016,steidelKeckLymanContinuum2018}. A complete theoretical and empirical understanding of the neutral-phase ISM and CGM, particularly how they affect emergent \lya{} and LyC across redshift, remains incomplete.
	
	Using the promising results found at $z\sim0.3$, inferences have been made about escape fractions of galaxies at $z\gtrsim2$ with high-resolution spectroscopy of Ly$\alpha$. In \citet{naiduSynchronyProductionEscape2022}, the authors used the Ly$\alpha$ profile shapes of a sample of 35 $z\sim2$ Ly$\alpha$ emitters (LAEs) observed with X-SHOOTER to infer their escape fractions. In \citet{mattheeReSolvingReionization2022}, the authors extended their framework to construct a model of reionization that progresses rapidly and ends late, and is primarily driven by LAEs. These two works are built upon the connection between Ly$\alpha$ profile shape and LyC escape, which remains untested beyond $z\sim0.3$. The $z\sim2$ galaxies within the X-SHOOTER survey did not have LyC measurements necessary to directly constrain \fesc{}, and thus definitively establish if \fesc{}-\vsep{} relationships are invariant with redshift.
	
	The Keck Lyman Continuum Survey (KLCS) represents an ideal opportunity for directly constraining the connection between \fesc{} and Ly$\alpha$ profile shape at the highest redshifts possible \citep{steidelKeckLymanContinuum2018}. The spectroscopic sample, which covers the LyC wavelength region, is both deep and has been cleaned from foreground contamination \citep{pahlUncontaminatedMeasurementEscaping2021}, which is a significant concern plaguing LyC surveys at $z\sim2-4$ \citep{vanzellaDetectionIonizingRadiation2012}. 
	Given the relatively low resolution of the Keck/LRIS spectra in the KLCS ($R\sim1300$), double peaked profiles are not reliably recovered at low velocity separations. Other probes of \lya{} kinematics are more robust to lower resolution, such as
	the peak velocity offset from systemic, and are measurable for a significant subset of the KLCS. This measurement is correlated with \vsep{} in low-redshift samples and is more readily observed at $z\gtrsim6$ when the blue side of Ly$\alpha$ is entirely attenuated by the neutral-phase IGM. Other spectral features that trace the velocity structure of the ISM and CGM of the galaxies, such as low-ionization metal absorption lines, are also recoverable given the depth of the Keck/LRIS observations. A detailed analysis of the shape of Ly$\alpha$ profiles in the KLCS is therefore of great value to understanding whether the mechanisms of LyC escape are consistent across redshift, and for testing how the shape of Ly$\alpha$ correlates with \fesc{} at high redshift before \lya{}-based \fesc{} diagnostics are applied to galaxies in the epoch of reionization. This work serves as an extension to the discussion of \lya{} emission-line kinematics in the KLCS presented in \citet{steidelKeckLymanContinuum2018}, and features analyses of \lya{} profiles of individual galaxies, explicit stacks as a function of \lya{} profile shape, and comparison to recent low- and high-redshift surveys of galaxies with \lya{} in emission.
	
	The paper is organized as follows. In Section \ref{sec:methods}, we describe the KLCS sample and follow-up observations, lay out the methodology for making quantitative measurements of Ly$\alpha$ profile shapes, and review our spectral-stacking and fitting techniques. In Section \ref{sec:res}, we present the \lya{} measurements of the KLCS and the modeled \fesc{} of composite spectra binned across such measurements. In Section \ref{sec:disc}, we discuss the applicability of our results for reionization-era observations, and compare to findings to those at low redshift and the \fesc{} assumptions that have been applied to $z\sim2$ and beyond. In Section \ref{sec:summary}, we summarize the main results of this work.
	
	Throughout this paper, we adopt a standard $\Lambda$CDM cosmology with $\Omega_m$ = 0.3, $\Omega_{\Lambda}$ = 0.7 and $H_0$ = 70 $\textrm{km\,s}^{-1}\textrm{Mpc}^{-1}$. The {\fesc} values reported in this paper are absolute escape fractions, equivalent to {\fesca} in \citet{steidelKeckLymanContinuum2018}, and defined as the fraction of all H-ionizing photons produced within a  galaxy that escapes into the IGM. We also employ the AB magnitude system \citep{okeSecondaryStandardStars1983}.
	
	\section{Sample and Methodology} \label{sec:methods}
	
	In order to understand the connection between {\lya} profile shape and escaping ionizing radiation from galaxies at high redshift, we require both spectroscopy of Ly$\alpha$ with sufficient resolution to discern its structure and measurements of the Lyman Continuum (LyC) region to constrain \fesc{}. These analyses must be applied to statistical samples with adequate numbers of galaxies that allow for binning as a function of a given measurement of the Ly$\alpha$ profile, given the uncertain correction of the IGM on individual LyC flux density measurements. The galaxies in the KLCS have deep rest-UV spectroscopy that covers both \lya{} and the LyC, as well as low-ionization interstellar absorption that probes outflowing gas in the ISM+CGM, and rest-optical spectroscopy covering nebular emission lines useful for determining systemic redshifts. In this section, we summarize the observations available for the KLCS sample and describe our measurements that quantify kinematic information contained within the \lya{} profiles of KLCS Ly$\alpha$ emitters.
	
	\subsection{Uncontaminated KLCS}
	
	The KLCS is a deep, spectroscopic exploration of the escaping ionizing signal from a sample of 136 star-forming, Lyman break galaxies (LBGs) at {\ziii}. The survey was conducted using the Low Resolution Imaging Spectrometer \citep[LRIS, ][]{okeKeckLowResolutionImaging1995,steidelSurveyStarForming2004} on the Keck I telescope from 2006 to 2008. Each of the objects in the nine survey fields were observed for an average of 9 hours in order to sensitively probe the LyC spectral region. After removing objects with clear spectroscopic blends and instrumental defects, the final sample of 124 galaxies was presented in \citet{steidelKeckLymanContinuum2018}. Included in this parent sample were 15 galaxies individually detected in LyC, defined as $f_{900} > 3\sigma_{900}$, where $f_{900}$ is the average flux density $880 \leq \lambda_{0}/\angstrom \leq 910$ and $\sigma_{900}$ is the $f_{900}$ measurement uncertainty.
	Measurements of the LyC at high redshift require careful consideration of contamination from foreground sources, which can significantly bias average LyC fluxes. High-resolution imaging is necessary for resolving and analyzing foreground objects \citep{vanzellaDetectionIonizingRadiation2012,mostardiHIGHRESOLUTIONHUBBLESPACE2015}. The KLCS was the subject of a follow up by the \textit{Hubble Space Telescope} (\textit{HST}) in order to explore this contamination. Using {\vjh} imaging, we selected contaminates on the basis of their colors and removed them from the sample, reducing the sample size to 120 (with 13 individually detected in LyC) and reporting a revised average {\fesc} of $0.06\pm0.01$ for the KLCS \citep{pahlUncontaminatedMeasurementEscaping2021}.
	
	In addition to deep, rest-UV LRIS spectroscopy and \textit{HST} imaging, a wealth of ancillary data is available for the objects in the KLCS. Both ground-based and space-based photometry were analyzed in tandem with spectral stacking in order to explore {\fesc} and galaxy properties in \citet{pahlConnectionEscapeIonizing2023}. These properties, including stellar mass (\mstar{}), star-formation rate (SFR), E(B-V), and stellar age, were measured using stellar-population synthesis (SPS) models for 96 KLCS objects. We recovered inverse trends between \fesc{} and \mstar{} as well as \fesc{} and E(B-V), and no significant trends between \fesc{} and stellar mass or specific star-formation rate (sSFR).
	
	\subsection{Redshift measurements and velocity analysis} \label{sec:z_methods}
	
	Of the galaxies within the uncontaminated KLCS, 80 have Ly$\alpha$ profiles such that a redshift can be estimated from at least one peak of emission. Here, we analyze the number and position of peaks within the Ly$\alpha$ profiles of these 80 objects. We quantify the position of the peaks relative to systemic redshifts, which are based on rest-optical nebular emission lines, and low-ionization interstellar absorption redshifts. The velocity separations that we endeavor to measure, which include the \lya{} peak separation ($\vsep$), the separation between the red-side \lya{} peak and the systemic redshift (\vlyar{}), and the separation between the red-side \lya{} peak and the low-ionization interstellar redshift (\vlma{}), are illustrated for a KLCS galaxy in Figure \ref{fig:lyaspec2} as a demonstration of our methodology.
	
	\subsubsection{Ly$\alpha$ emission peaks}
	
\begin{figure*} %
		\centering
		\includegraphics[width=\textwidth]{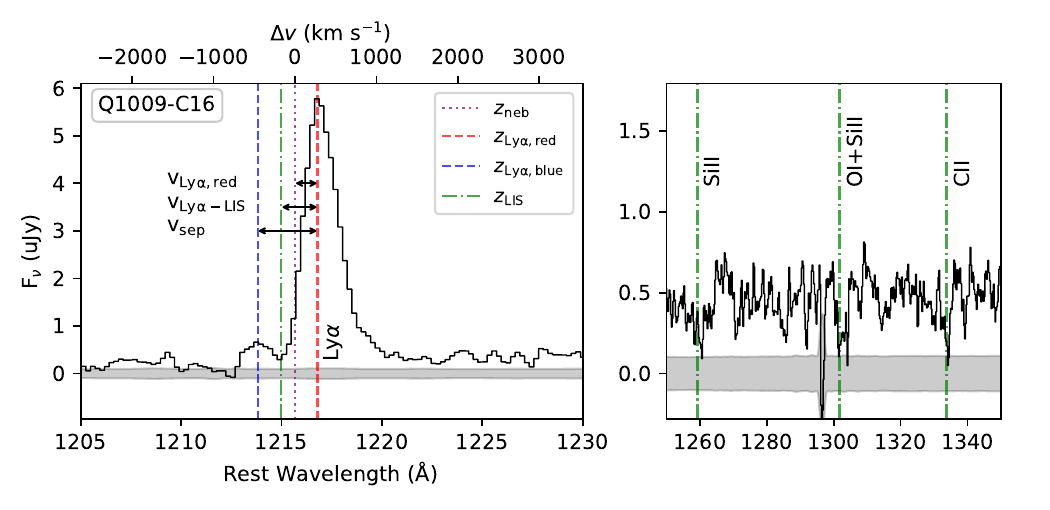}
		\caption{Ly$\alpha$ velocity measurements for the KLCS galaxy Q1009-C16 overlaid on its Keck/LRIS spectrum. The velocities are calculated as the difference between different redshift estimates: \vlyar{} is the difference between \zlyar{} and \zsys{}, \vlma{} is the difference between \zlyar{} and \zlis{}, and \vsep{} is the difference between \zlyar{} and \zlyab{}.
			\textbf{Left panel:} The Ly$\alpha$ profile of Q1009-C16. $\zlyar=3.1662$ and $\zlyab=3.1557$ are determined from the wavelength corresponding to the maximum flux density within the red- and blue-side Ly$\alpha$ profile, respectively. $\zsys=3.1621$ is determined from rest-optical nebular emission lines.
			\textbf{Right panel:} Three low-ionization metal absorption lines detected in Q1009-C16, which are used to determine $\zlis=3.1607$.
		}
		\label{fig:lyaspec2}
	\end{figure*}

	The velocity separation of the blue- and red-side peaks of Ly$\alpha$ emission (\vsep{}) has been demonstrated to strongly correlate with \fesc{} at low redshift \citep{verhammeUsingLymanDetect2015,izotovLowredshiftLymanContinuum2018}, indicating that \vsep{} increases with optical depth in \lya{}, H~\textsc{i} column density, and/or neutral-gas covering fraction. Of the 80 KLCS objects with \lya{} emission, 24 have visual evidence of double-peak structure within their 1D spectra. These \lya{} profiles were typically red-side dominated: we show the spectrum of a KLCS object (Q0100-C16) with double-peaked structure as an example in Figure \ref{fig:lyaspec2}. Of these 24 objects, we identified 15 with sufficiently-resolved blue-side peaks, such that the wavelength of peak emission could be estimated for both sides of the bimodal emission profile. In order to make an estimate the peak wavelength for the blue-side emission, we first fit a Gaussian profile and defined a window that lies within 1$\sigma$ of the central wavelength.
	We then determined $\lambda_{\rm Ly\alpha,blue}$, which corresponded to the wavelength in which maximum flux occurred within this window. This procedure ensured that any asymmetries within the blue-side profile were accounted for. We defined the redshift of the peak of blue-side \lya{} emission as:
	\begin{equation}
		\zlyab =\frac{\lambda_{\rm Ly\alpha,blue}}{1215.67\AA} - 1.
	\end{equation}
	
	We determined the redshift corresponding to the peak of red-side Ly$\alpha$ emission in an identical manner to that of the blue side, save for four objects. For these four profiles, the \lya{} emission was either broad or noisy, and did not have a well-defined maximum flux density. We thus calculated \zlyar{} using simply the central wavelength of the fitted profile. 
	
	We define \zlyar{} as 
	\begin{equation}
		\zlyar =\frac{\lambda_{\rm Ly\alpha,red}}{1215.67\AA} - 1,
	\end{equation}
	where $\lambda_{\rm Ly\alpha,red}$ is either the wavelength corresponding to the maximum flux within 1$\sigma$ of the central wavelength of the best-fit Gaussian, or the central wavelength itself. In Figure \ref{fig:lyaspec2}, we display \zlyar{} for Q0100-C16 as a red, dashed line and \zlyab{} as a blue, dashed line. We note that we measured only one \lya{} redshift for the majority (65/80) of the KLCS objects, considering these had no resolved double-peaked structure. We define these singular redshifts as \zlyar{} for simplicity, but add no requirement that these peaks be necessarily redshifted.
	
	We estimated the errors on both \zlyar{} and \zlyab{} measurements in a similar fashion. After fitting a Gaussian to a given profile to determine the window in which the peak flux would be computed, we perturbed the observed spectrum according to the error spectrum. Each flux density value at a given wavelength was randomly modulated according to a Gaussian with zero mean and width equal to the corresponding error spectrum value. From this perturbed spectrum, we recorded the wavelength corresponding to the maximum flux within 1$\sigma$ of the central wavelength of the initial best-fit Gaussian. Each spectrum was perturbed 100 times in this manner, resulting in 100 \zlyar{} (\zlyab{}) measurements for each object. The final \zlyar{} (\zlyab{}) estimate was taken as the mean of these 100 values, with the error on \zlyar{} (\zlyab{}) taken as the standard deviation of the 100 values. For the four objects in which $\lambda_{\rm Ly\alpha,red}$ was estimated from the best-fit Gaussian alone, we performed a similar procedure, but instead re-fit the Gaussian using the perturbed spectrum, resulting in 100 values for \zlyar{} estimated from the central wavelength of each best-fit Gaussian.
	
	\subsubsection{Systemic redshifts}
	
	We determined the systemic redshift ($z_{\text{sys}}$) from rest-optical nebular emission lines. From our parent sample of 80 objects with Ly$\alpha$ emission, 45 objects have either \oiii{}, \oii{} or \hb{} emission in existing Keck/MOSFIRE spectroscopy. We fit Gaussian profiles to these lines, and used the central wavelength of well-fit profiles to calculate the object's systemic redshift. If multiple of these lines were recovered in the rest-optical spectrum, we estimated {\zsys} by averaging the redshifts implied by each line measurement. The nebular redshift for Q1009-C16 is presented as a dotted, purple line in Figure \ref{fig:lyaspec2}. Given the strength and narrow width of the nebular emission lines in these objects, in addition to the higher spectral resolution and SNR of the Keck/MOSFIRE spectra, we consider the error on \zsys{} negligible compared to redshifts measured from Keck/LRIS.
	
	\subsubsection{Low-ionization absorption redshifts}
	
	The wavelength position of interstellar metal absorption lines traces the velocity of low-ionization material in the ISM/CGM, which tends to be negative (outflowing) at these redshifts \citep{shapleyRestFrameUltravioletSpectra2003,steidelStructureKinematicsCircumgalactic2010,jonesKeckSpectroscopyFaint2012,duRedshiftEvolutionRestUV2018}. We determined the low-ionization absorption redshift by using the same Keck/LRIS spectra used for \lya{} analysis, as the spectral coverage extends to $\sim1600$\AA{} in the rest frame. The strongest low-ionization absorption features in these spectra were typically 
	Si~\textsc{ii}$\lambda1260$, O~\textsc{i}$\lambda1302+$Si~\textsc{ii}$\lambda1304$, and C~\textsc{ii}$\lambda1334$. We estimated $z_{\rm LIS}$ by fitting Gaussian profiles to these three lines, and averaging the redshifts calculated from the central wavelengths of well-constrained Gaussian fits. 
	For O~\textsc{i}$\lambda1302+$Si~\textsc{ii}$\lambda1304$, which is blended at the resolution of our spectra, we assumed a singular wavelength of 1302.5\AA.
	In the right panel of Figure \ref{fig:lyaspec2}, we display Si~\textsc{ii}$\lambda1260$, O~\textsc{i}$\lambda1302+$Si~\textsc{ii}$\lambda1304$, and C~\textsc{ii}$\lambda1334$ for Q1009-C16. The $z_{\rm LIS}$ measured from these three lines is displayed as a green, dash-dot line. The same redshift is also displayed in the left panel, showing its position relative to Ly$\alpha$.
	
	We estimated the errors on \zlis{} empirically, again by using the Keck/LRIS error spectra. We performed Monte Carlo simulations by perturbed each flux density value by a Gaussian with zero mean and width equal to the value of the error spectrum at the corresponding wavelength. We then re-fit Gaussian profiles to each line using this perturbed spectrum, and re-calculated \zlis{}. For each object, we performed this perturbation 100 times, and measure 100 values for \zlis{}. The subsequent \zlis{} measurement and error was determined as the mean and standard deviation of these 100 values.
	
	\subsubsection{Velocities}
	
	Different kinematic measurements are possible for subsets of the KLCS, depending on which spectral features were detected. For the objects for which we measured \zlyab{}, \zlyar{}, and \zsys{}, we define the separation between the blue- and red-side \lya{} peak as
	\begin{equation}
		\vsep = c \times \frac{\zlyar - \zlyab}{1 + \zsys}.
	\end{equation}
	For objects with only \zlyab{} and \zlyar{}, we use an average of \zlyab{} and \zlyar{} rather than \zsys{} to define relative velocities:
	\begin{equation}
		\vsep = c \times \frac{\zlyar - \zlyab}{1+(\zlyar + \zlyab)/2}.
	\end{equation}
	We defined the \ksep{} sample as the 15 objects with this measurement. Of these 15, only one is individually-detected in LyC.
	
	While \vsep{} has been demonstrated to be an \fesc{} diagnostic in low-redshift ($z\sim0.3$) LyC surveys, its observability becomes increasingly difficult at higher redshifts as the blue-side peak becomes attenuated by the neutral-phase IGM. We introduce \vlyar{} as the velocity separation between the red-side \lya{} peak and systemic redshift, as a \vsep{}-like quantity more readily measured both in the KLCS and in reionization-era galaxies. We measured this quantity for the galaxies with \zlyar{} and \zsys{} as
	
	\begin{equation}
		\vlyar = c \times \frac{\zlyar - \zsys}{1 + \zsys},
	\end{equation}
	and defined the \klya{} sample as the 45 objects with this measurement. Eight of the objects within \klya{} are LyC detections. We lack Keck/MOSFIRE observations of the full KLCS, as objects at $2.75\lesssim z\lesssim 2.95$ lack strong nebular lines in atmospheric windows. We therefore investigate the velocity separation of \zlyar{} and \zlis{}, which contrasts the position of \lya{} to the velocities of neutral ISM. We define this quantity as
	
	\begin{equation}
		\vlma = c \times \frac{\zlyar - \zlis}{1+(\zlyar + \zlis)/2},
	\end{equation}
	and defined the \klma{} sample as the 67 objects with both \lya{} emission and LIS absorption redshifts measured. This sample includes six of the 13 LyC detections in the full KLCS.
	We overlay \vsep, \vlyar, \vlma{} on the Ly$\alpha$ profile of Q1009-C16 in Figure \ref{fig:lyaspec2}. Due to its complete suite of redshifts measurements, this object was included in all three KLCS subsamples.
	
\begin{figure*} %
		\centering
		\includegraphics[width=\textwidth]{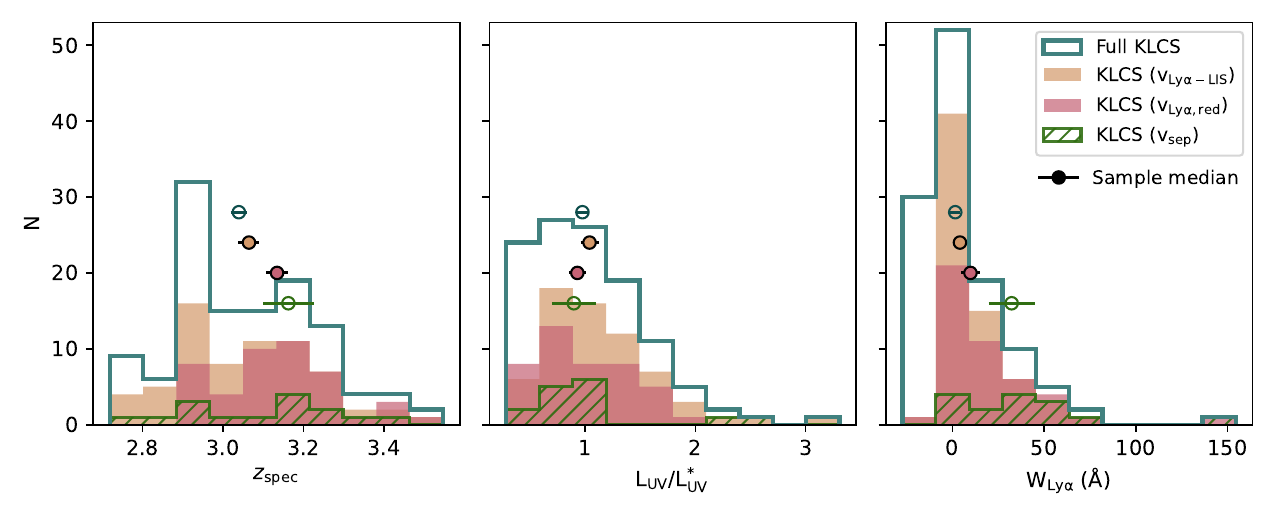}
		\caption{Distributions of $z_{\rm spec}$, \luv{}/$L_{\rm UV}^*$, and \wlya{} for the full KLCS as well as the \klma, \klya, and \ksep{} subsamples. The full KLCS includes all 120 galaxies as defined in \citet{pahlUncontaminatedMeasurementEscaping2021}. The \klma{} sample contains the 68 objects with \zlyar{} and \zlis{} measurements, while \klya{} contains the 45 objects with \zlyar{} and \zsys{} measurements. The \ksep{} sample includes the 15 objects with resolved double-peaked \lya{} structure. The medians of the four samples with respect to $z_{\rm spec}$, \luv{}, and \wlya{} are displayed as circles, with the respective standard errors on the median displayed as horizontal error bars. In vertical descending order, the circles correspond to: the full KLCS, \klma{}, \klya{}, and \ksep{}. The median \wlya{} and \luv{} of the \klma{} and \klya{} subsamples are consistent with those the full KLCS, while \ksep{} features a larger median \wlya{}.}
		\label{fig:sample}
	\end{figure*}

	We summarize the combination of detected emission lines required for each of the three KLCS subsamples, including their resulting sample sizes, in Table \ref{tab:samp}.
	We additionally display sample statistics of the full KLCS alongside the three KLCS subsamples in Figure \ref{fig:sample}. We include the distributions of spectroscopic redshifts, UV luminosities, and \lya{} equivalent widths (\wlya{}) for each subsample, alongside their respective medians and standard errors of the median. The \klya{} and \klma{} samples have median \wlya{} and UV luminosity consistent with those of the full KLCS, within one standard error. The \ksep{} sample has a significantly-higher median \wlya{} than that of the full KLCS, demonstrating that \vsep{} measurements require larger equivalent widths, such that the flux blueward of systemic was sufficiently detected.
	
	\begin{deluxetable*}{llc} \label{tab:samp}
		\tablecaption{Emission line criteria for the KLCS subsamples.}
		\tablehead{& \colhead{N} & \colhead{Required lines}}
		\startdata
			\ksep{} & 15 & \lya{} with double peaked structure \\
			\klya & 45 & \lya; \oiii{}, \oii{}, or \hb{} \\
			\klma & 69 & \lya{}; Si~\textsc{ii}$\lambda1260$, O~\textsc{i}$\lambda1302$+Si~\textsc{ii}$\lambda1304$, or C~\textsc{ii}$\lambda1334$\\
		\enddata
	\end{deluxetable*}
	
	\subsection{Sample binning and spectral modeling} \label{sec:spec}
	
	While $f_{900}$ can be measured for each object individually, constraining the LyC leaking in the vicinity of a galaxy requires an understanding of the attenuation of the signal from neutral hydrogen along the line of sight in the IGM and CGM. The transmission of LyC emission varies significantly from sightline to sightline at the redshifts of our sample, introducing large uncertainties on individual LyC measurements \citep{rudieColumnDensityDistribution2013,steidelKeckLymanContinuum2018}. To circumvent this sightline to sightline variability, we used binned subsamples and composite spectra that reflect average effects of IGM and CGM attenuation on the LyC spectral region, as in \citet{steidelKeckLymanContinuum2018}. 
	Given the range in sample sizes of the three KLCS subsamples, we discuss the feasibility of recovering trends between \fesc{} and \lya{} profile shape in Appendix \ref{sec:app2}, considering the inherent uncertainties introduced by sample variance and IGM/CGM variability.
	We excluded \ksep{} from any \fesc{} analyses due to its small sample size and the large $f_{900}$ uncertainties associated with binned subsamples with size $\leq8$. We included \klya{} and \klma{} and considered them appropriate for investigating \fesc{} vs. \vlyar{} and \fesc{} vs. \vlma{}, respectively.
	
	In order to understand how ionizing-spectral properties vary with \vlyar{} and \vlma{}, we bin the \klya{} sample as a function of \vlyar{} and the \klma{} sample as a function of \vlma{}. We create two bins for each property, with each \vlyar{} bin containing 22 or 23 galaxies, and each \vlma{} bin containing 33 or 34 galaxies\footnote{We assign the extra galaxy to the bin containing objects with larger velocity measurements.}.
	We additionally bin both \klya{} and \klma{} by \wlya{}.
	
	For each subsample, we generated composite spectra representing the average spectral properties of the component galaxies. We follow the methodology of \citet{steidelKeckLymanContinuum2018} \citep[also see][]{pahlUncontaminatedMeasurementEscaping2021,pahlSearchingConnectionIonizingphoton2022,pahlConnectionEscapeIonizing2023}. Briefly, each individual spectrum is first normalized to the average flux density in the non-ionizing UV spectral region. We then computed the sigma-clipped mean of the distribution of flux densities at each rest-frame wavelength increment, with clipping applied at 3$\sigma$. We did not apply sigma clipping to the Ly$\alpha$ spectral region ($1200-1230$\AA{}) to avoid censoring the actual line emission in this region.
	
	We determined \fobs{}, a metric defined as the ratio between the mean flux densities in the Lyman continuum (LyC) region ($880-910$\AA{}, $f_{900}$) and the non-ionizing UV continuum ($1475-1525$\AA{}, $f_{1500}$), for each composite spectrum. While this ratio offers insights into the average observed leakage of ionizing photons, it is necessary to address the impact of reduced transmission from the IGM in the LyC region in order to understand the emergent ionizing signal at the galaxy's edge.
	We applied corrections to the spectra using the average "IGM+CGM" transmission functions sourced from \citet{steidelKeckLymanContinuum2018}. These corrections were computed at the mean redshift specific to each composite subsample. The basis for these corrections was grounded in the statistics of H~\textsc{i} absorption systems along QSO sightlines, as elucidated by \citet{rudieGaseousEnvironmentHighz2012} and \citet{rudieColumnDensityDistribution2013}.
	After applying the correction, we recalculated the ratio of $f_{900}$ to $f_{1500}$ to derive \fout{}, introduced earlier. This metric encapsulates the ratio that would be observed at $50$ proper kpc from the galaxy center, as detailed in \citet{steidelKeckLymanContinuum2018}.
	
	For each composite spectrum, we derive \fesc{} by again following the methodology originally presented in \citet{steidelKeckLymanContinuum2018}. While \fout{} is a useful empirical metric for quantifying leaking LyC radiation and does not rely on model choices, \fesc{} retains widespread use in reionization models. 
	We used the Binary Population and Spectral Synthesis stellar-population synthesis models \citep[BPASS v2.2.1][]{eldridgeBinaryPopulationSpectral2017}, combining these models with a SMC extinction curve \citep{gordonQuantitativeComparisonSmall2003}, and considered a range of E(B-V) values from 0.0 to 0.6, along with a fixed metallicity of 0.07 times solar.
	To model the ISM, we adopt the ``holes" model, assuming that LyC light escapes through a patchy neutral-phase gas \citep{zackrissonSpectralEvolutionFirst2013,reddyConnectionReddeningGas2016,reddyEffectsStellarPopulation2022,steidelKeckLymanContinuum2018}. The free parameters of the fit are the neutral gas covering fraction $f_{\rm c}$, the column density of neutral hydrogen N$_{\rm HI}$, and the dust attenuation from the foreground gas E(B-V)$_{\rm cov}$ (with the uncorrected portion assumed to be dust-free). In the ``holes" model, \fesc{} is derived from $f_{\rm c}$, where $\fesc=1-f_{c}$.
	
	In order to estimate the uncertainty in average escape parameters for a given set of galaxies, we must understand the level of variability induced from sample construction and from the errors on individual galaxy properties. We follow the procedure described in \citet{pahlConnectionEscapeIonizing2023}. We randomly perturb each measurement (\vlyar{}, \vlma{}, \wlya{}) by a Gaussian with width equal to its error. We sorted objects into two bins using the new, perturbed parameter histograms. In order to incorporate sample variance, we then bootstrap resampled (with replacement) the galaxies in each binned sample. We performed this procedure 100 times, in order to generate 100 sets of objects for each binned subsample.
	Following that, we generated composite spectra and assessed the escape of ionizing photons for each random draw, employing the previously outlined procedure in this section. The average and standard deviation of the distributions of \fobs{}, \fout{}, and \fesc{} derived from the 100 composite spectra served as the fiducial value and error estimate for the respective binned sample.

	\section{Results} \label{sec:res}
	
\begin{figure*} %
		\centering
		\includegraphics[width=0.8\textwidth]{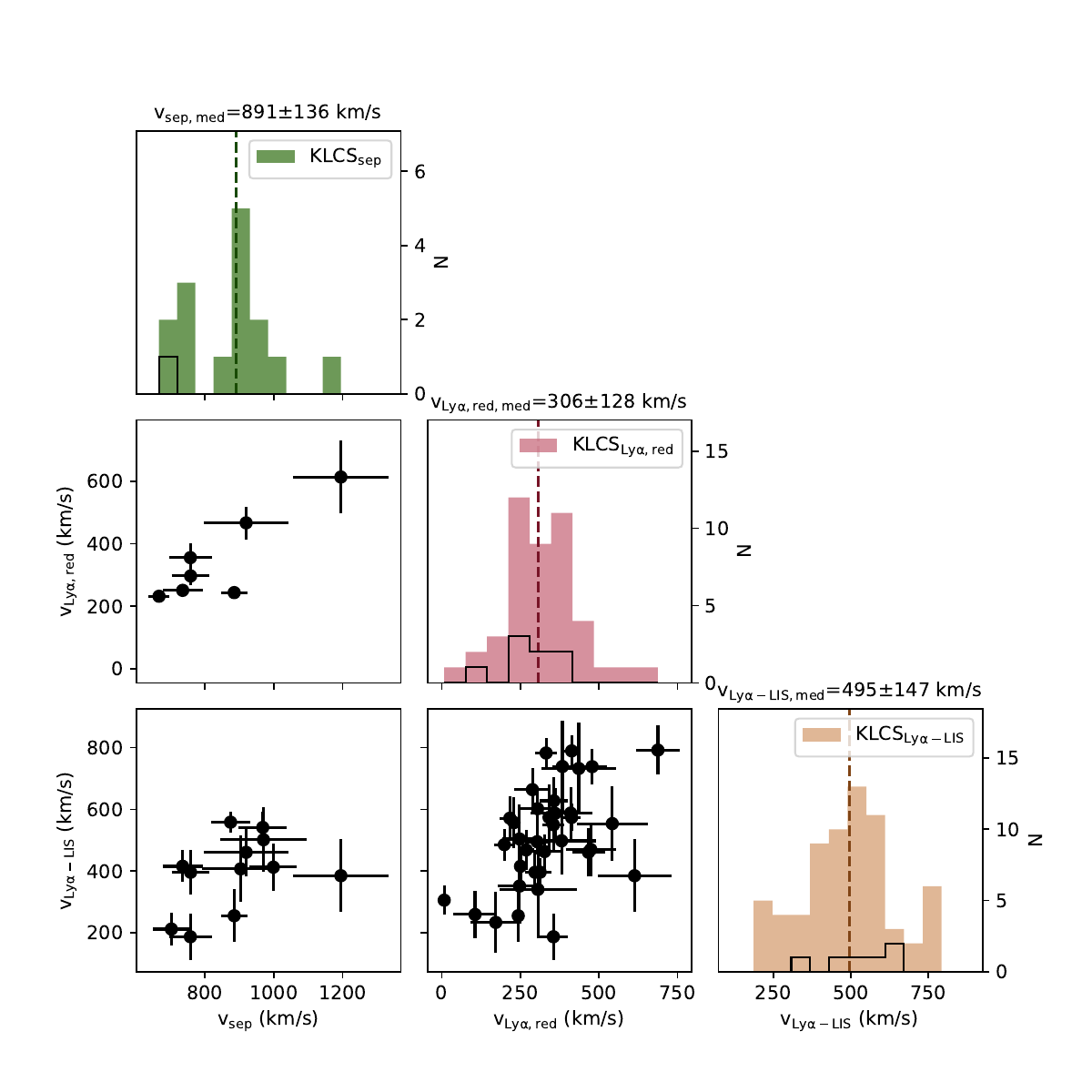}
		\caption{Corner plot of \lya{} velocity measurements for the KLCS. 
			\textbf{Upper-right panels:} Histograms of \vsep{} in green, \vlyar{} in red, and \vlma{} in yellow. The median of each distribution is displayed as a dashed line. For \vlyar{} and \vlma{}, this median defines the edge of the two bins used for composite spectrum generation for \vlyar{}. Medians and standard deviations are also displayed above each histogram. Objects individually detected in LyC are overplotted as transparent, black histograms.
			\textbf{Lower-left panels:} Relationships between \vsep{}, \vlyar{}, and \vlma{}. Only objects with both measurements are displayed. Typical errors are shown in black in the lower right of each panel. We find a significant (p$\sim0.001$) correlation between \vlma{} and \vlyar{}. We also see a correlation between \vlyar{} and \vsep{}, but the sample size is insufficient for formal analysis.
		}
		\label{fig:vcompare}
	\end{figure*}

	In order to understand how \fesc{} depends on \lya{} profile shape and kinematics, we measured \vsep{}, \vlyar{}, and \vlma{} for different subsets of the KLCS. We display these velocity measurements in the histograms in Figure \ref{fig:vcompare}. Within each histogram, the median value is displayed as a dashed, vertical line. This line represents the bin edge from which  two equally-sized samples were constructed for composite spectra generation. We also display the relationships between the velocity measurements in the scatterplots in Figure \ref{fig:vcompare}. We see correlations between \vlyar{} and \vsep{}, and no correlation between \vlma{} and \vsep{} although the sample of KLCS galaxies with \vsep{} measurements is too small to do formal analysis. The sample of objects with both \vlyar{} and \vlma{} measurements is 36, large enough to use the Spearman rank correlation test to quantify the relationship between the two variables. With the null hypothesis being that \vlyar{} and \vlma{} are uncorrelated, we find a correlation coefficient of 0.487 with a p-value of 0.003. This result significantly rejects the null hypothesis, and indicates that \vlyar{} is correlated with \vlma{}.
	
\begin{figure}[hbt!] %
		\centering
		\includegraphics[width=\columnwidth]{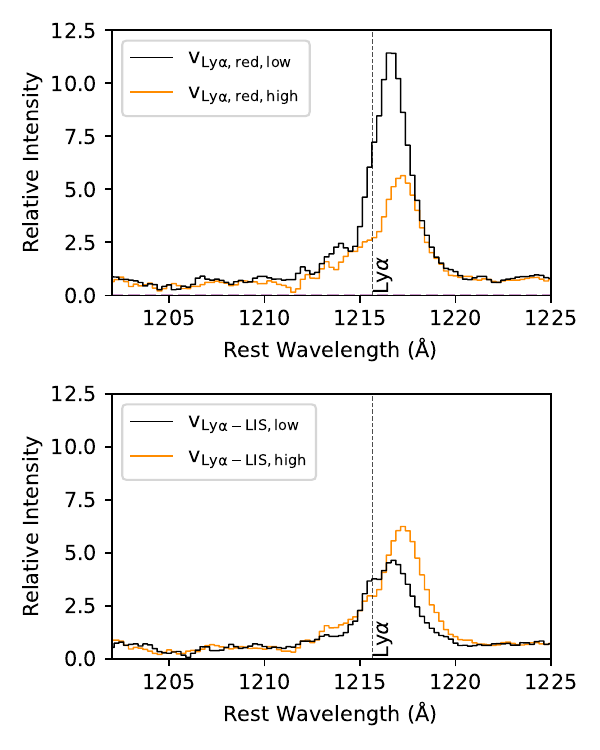}
		\caption{Composite KLCS spectra binned as a function of different spectral parameters. The composite constructed from objects with lower values is displayed as a black curve, while the objects with the highest values are averaged into a composite shown with an orange curve. The upper panel features two bins of \vlyar{}, with the lower bin containing 22 galaxies while the upper bin contained 23. The lower panel is instead binned by \vlma{}, with the lower bin containing 33 galaxies and the upper bin containing 34. Both \vlyarl{} and \vlmal{} feature more flux at systemic than their counterparts, and the \vlyarl{} composite has a larger \wlya{} than the \vlyarh{} composite.
		}
		\label{fig:binspec}
	\end{figure}

	For each binned sample illustrated in Figure \ref{fig:vcompare}, we created a composite in order to understand their average ionizing and non-ionizing rest-UV spectra. To highlight the \lya{} profiles of the objects in each bin, we display composites generated from bins of \vlyar{} and \vlma{} in Figure \ref{fig:binspec}. Both \vlyarh{} and \vlmah{} feature \lya{} profiles shifted to larger redshifted velocities. Meanwhile, \vlyarl{} and \vlmal{} have a larger amount of \lya{} flux observed at systemic, which indicates objects with these properties have more regions in their ISM and CGM with low-column-density ($\lesssim10^{12}$\:cm$^{-2}$), neutral-phase gas. Double peaked structure is apparent in all composites. However, all composites save for \vlyarl{} have a blue-side profile that is unresolved from the stronger, red-side profile. Finally, \vlyarl{} features stronger \lya{} emission than \vlyarh{}, indicating that galaxies with \lya{} profiles less offset from systemic also have larger \lya{} equivalent widths, as both are modulated by the covering fraction of neutral gas. Consistent with this result, surveys of LAEs have been demonstrated to have lower \vlyar{} (and higher \wlya{}) than typical LBGs \citep[e.g.,][]{trainorSpectroscopicPropertiesLyaEmitters2015}. In contrast with the KLCS \vlma{} composites, surveys of LBGs have shown a direct correlation between \vlma{} and \wlya{}, although across a larger dynamic range of \vlma{} than is probed by our bins \citep[e.g.,][]{shapleyRestFrameUltravioletSpectra2003}.

\begin{figure*} %
		\centering
		\includegraphics[width=0.65\textwidth]{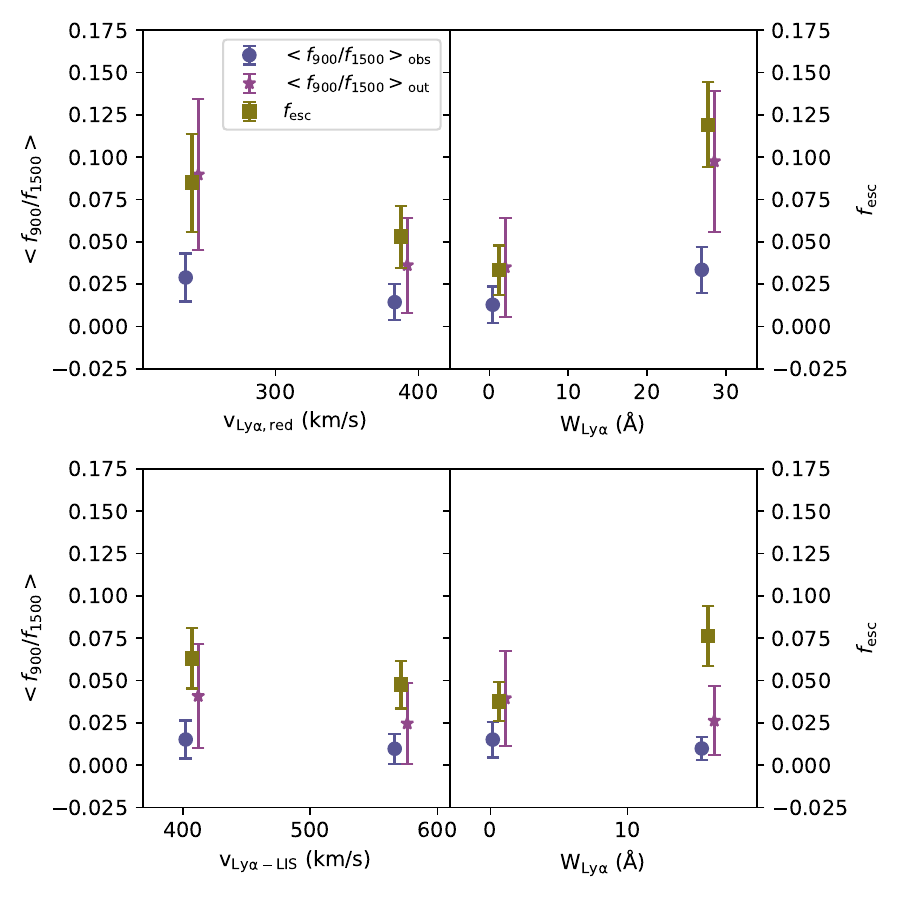}
		\caption{Different measurements of ionizing photon escape estimated for binned subsamples of the KLCS. Each measurement is displayed with respect to the median property of the objects contained in each bin. The blue circles are \fobs{} measurements, which are computed directly from the composite spectra. The purple stars are \fout{} measurements, which include a correction to $f_{900}$ from attenuation from the IGM and outer CGM. Estimates of \fesc{} of composite spectra are displayed as gold squares, which rely on stellar-population synthesis modeling and ISM geometry featuring ``holes" free from neutral gas and dust \citep{steidelKeckLymanContinuum2018,reddyConnectionReddeningGas2016,reddyEffectsStellarPopulation2022}. The blue circles and purple stars are slightly shifted left and right, respectively, for visual clarity. The upper two panels are both constructed from the same parent sample, which are all KLCS objects with both \lya{} and systemic velocity measurements. The lower two panels are constructed from a sample of galaxies with \lya{} emission and LIS absorption velocity measurements. We measure a significant difference in \fout{} across the two bins of \vlyar{}, while we find no significant difference in either \fout{} or \fesc{} across bins of \vlma{}.
		}
		\label{fig:fesc}
	\end{figure*}

	From these composites, we conducted three assessments of ionizing photon escape, as explained in Section \ref{sec:spec}. The first measurement, denoted as \fobs{}, represents the ratio of ionizing to non-ionizing flux density observed directly in the composite derived from individual spectra. The second measurement, \fout{}, is the same ratio calculated from a composite spectrum that has been adjusted for the average line-of-sight attenuation caused by the IGM and CGM. Lastly, \fesc{} is a parameter estimated through stellar population synthesis modeling of the entire rest-UV composite. In Figure \ref{fig:fesc}, we present these three assessments of ionizing escape along with their respective uncertainties, organized by galaxy property for each subsample. The numerical values corresponding to these results can be found in Table \ref{tab:fit}, which also includes the median properties of the galaxies within each subsample.
	
	In order to determine whether ionizing-photon escape is correlated with a given kinematic measurement of \lya{}, we determine significance as defined by \citet{pahlSearchingConnectionIonizingphoton2022}. For two composites constructed via binning across a galaxy property, \fesc{} is significantly correlated with this property only when
	
	\begin{equation} \label{eqn:sig}
		|f_{\rm esc,high}-f_{\rm esc,low}|>\sqrt{(\sigma_{f_{\rm esc,high}})^2 + (\sigma_{f_{\rm esc,low}})^2},
	\end{equation}
	where $f_{\rm esc,highest}$ is measured from the bin containing objects with the largest of a binned value, and $f_{\rm esc,lowest}$ is measured from the bin containing objects with the smallest values. Significance between \fout{} and \fobs{} and galaxy property are determined in an identical fashion.
	
	Both \fesc{} and \fout{} appear to decrease at greater \vlyar{} in the upper-left panel of Figure \ref{fig:fesc}. According to Equation \ref{eqn:sig}, the difference in \fout{} measurements between the two bins of \vlyar{} is significant (1$\sigma$). The difference between \fesc{} measurements, however, does not meet our significance threshold. For a given composite, \fout{} may not precisely trace \fesc{}, as \fesc{} depends on stellar population modeling and fits to the entire rest-UV spectrum, including Lyman-series absorption lines.
	The trends with \vlyar{} can be compared to correlations between \fout{}, \fesc{}, and \wlya{} within the same subsample, shown in the upper right panel. 
	We see a significant difference in both \fout{} and \fesc{} (1$\sigma$ and 2$\sigma$, respectively) between the two bins of \wlya{} in the \klya{} subsample, reflecting the correlation of the two parameters in the full KLCS \citep{steidelKeckLymanContinuum2018,pahlUncontaminatedMeasurementEscaping2021}. Despite any differences in the ionizing spectra across bins of \vlyar{}, it is clear that any anti correlation between \fesc{} and \vlyar{} is less significant than that between \fesc{} and \wlya{}. Nonetheless, the fact that significant correlations are found when binning across \vlyar{} and \wlya{} is not surprising considering both lower \vlyar{} and higher \wlya{} probe conditions in the ISM and CGM that are conducive to ionizing-photon escape: lower covering fractions of neutral gas, which attenuates and resonantly-scatters Ly$\alpha$. 
	
	In contrast to \vlyar{}, in the the lower right panel of Figure \ref{fig:fesc}, we do not see a significant difference in \fesc{}, \fout{} or \fobs{} across the two bins of \vlma{}. This lack of significant difference can be compared to the significant difference in \fesc{} measured across two bins of \wlya{} in the \klma{} sample, shown in the lower right panel. This result indicates that \vlma{} may not be a robust proxy for \fesc{} at $z\sim3$. While \vlyar{} represents the amount that the peak of Ly$\alpha$ emission is offset from systemic due to resonant scattering from neutral-phase gas, \vlma{} includes the velocity offset from low-ionization absorption, which is sensitive to the velocity of the (typically) outflowing material. 
	Additionally, the metal-enriched gas responsible for LIS absorption may have a different velocity distribution and/or covering fraction than that of the gas that is responsible for the attenuation of \lya{} and LyC photons. Studies have shown that covering fractions derived from metal lines are systematically lower than those deduced from H~\textsc{i} absorption lines \citep[e.g.,][]{reddyConnectionReddeningGas2016,gazagnesNeutralGasProperties2018,reddyEffectsStellarPopulation2022}.
	We note that the bins in the \klma{} sample (lower panels) contain $\sim34$ galaxies each, while the bins in the \klya{} sample (upper panels) contain $\sim23$ galaxies. This difference introduces larger uncertainties in \fesc{} for the composites generated from bins of \vlyar{}, both from sample variance and errors on assumed IGM transmission fractions. However, \vlma{} measurements have typically-larger uncertainties than \vlyar{} (79\:km\:s$^{-1}$ and 39\:km\:s$^{-1}$, respectively). Thus, through our bootstrapping procedure, more galaxies near the \vlma{} bin edge will move across bins when generating perturbed samples, which will result in larger error bars in \fesc{}.
	
	\begin{deluxetable*}{lllllllll}[htbp!] \label{tab:fit}
		\tablecaption{Measurements of ionizing-photon escape for the \klya{} and \klma{} samples, binned according \lya{} strength and kinematics.}
		\tablehead{& \colhead{N} & \colhead{\thead{v$_{\rm Ly\alpha,red,med}$$^{a,b}$\\(km\:s$^{-1}$)}} & \colhead{\thead{v$_{\rm Ly\alpha-LIS,med}$$^{a,b}$\\(km\:s$^{-1}$)}} & \colhead{\thead{W$_{\rm Ly\alpha,med}$$^a$\\(\AA{})}} & \colhead{\thead{L$_{\rm UV,med}$/\\\quad L$_{\rm UV}^{*}$$^a$}} & \colhead{\thead[l]{$<f_{900}/$\\$f_{1500}>_{\rm obs}$$^a$}} & \colhead{\thead[l]{$<f_{900}/$\\$f_{1500}>_{\rm out}$$^a$}} & \colhead{$f_{\rm esc}$$^a$}}
		\startdata
		\multicolumn{9}{c}{\klya{}}\\
		\midrule
		all                         &  45 &                                     $323\pm13$ &                                            --- &                                 $9.8\pm3.7$ &                                   $0.93\pm0.09$ &                    $0.022\pm0.007$ &                    $0.061\pm0.021$ &     $0.06\pm0.01$ \\
		v$_{\rm Ly\alpha,red,low}$  &  22 &                                \bm{$242\pm14$} &                                            --- &                                $21.0\pm6.8$ &                                   $0.84\pm0.12$ &                    $0.029\pm0.014$ &                    $0.090\pm0.045$ &     $0.08\pm0.03$ \\
		v$_{\rm Ly\alpha,red,high}$ &  23 &                                \bm{$388\pm19$} &                                            --- &                                 $4.4\pm3.0$ &                                   $1.11\pm0.16$ &                    $0.014\pm0.011$ &                    $0.036\pm0.028$ &     $0.05\pm0.02$ \\
		W$_{\rm Ly\alpha,low}$      &  22 &                                     $362\pm34$ &                                            --- &                            \bm{$1.2\pm1.2$} &                                   $1.29\pm0.10$ &                    $0.013\pm0.011$ &                    $0.035\pm0.029$ &     $0.03\pm0.01$ \\
		W$_{\rm Ly\alpha,high}$     &  23 &                                     $255\pm29$ &                                            --- &                           \bm{$27.7\pm3.5$} &                                   $0.70\pm0.06$ &                    $0.033\pm0.013$ &                    $0.098\pm0.042$ &     $0.12\pm0.03$ \\
		\midrule
		\multicolumn{9}{c}{\klma{}} \\
		\midrule
		all                         &  68 &                                            --- &                                     $533\pm23$ &                                 $4.1\pm2.0$ &                                   $1.03\pm0.06$ &                    $0.015\pm0.006$ &                    $0.042\pm0.017$ &     $0.06\pm0.01$ \\
		v$_{\rm Ly\alpha-LIS,low}$  &  33 &                                            --- &                                \bm{$407\pm21$} &                                 $4.1\pm4.2$ &                                   $1.07\pm0.10$ &                    $0.015\pm0.011$ &                    $0.041\pm0.031$ &     $0.06\pm0.02$ \\
		v$_{\rm Ly\alpha-LIS,high}$ &  34 &                                            --- &                                \bm{$571\pm16$} &                                 $3.6\pm2.1$ &                                   $1.03\pm0.13$ &                    $0.010\pm0.009$ &                    $0.025\pm0.024$ &     $0.05\pm0.01$ \\
		W$_{\rm Ly\alpha,low}$      &  33 &                                            --- &                                     $498\pm22$ &                            \bm{$0.6\pm0.9$} &                                   $1.25\pm0.15$ &                    $0.015\pm0.011$ &                    $0.039\pm0.028$ &     $0.04\pm0.01$ \\
		W$_{\rm Ly\alpha,high}$     &  34 &                                            --- &                                     $481\pm44$ &                           \bm{$15.9\pm4.0$} &                                   $0.99\pm0.08$ &                    $0.010\pm0.007$ &                    $0.026\pm0.020$ &     $0.08\pm0.02$ \\
		\enddata
		\tablenotetext{a}{The medians and standard error of the subsamples with respect to a given galaxy property, with the binned parameter highlighted in bold.}
		\tablenotetext{b}{Medians were excluded if measurements weren't available for all objects within a subsample.}
	\end{deluxetable*}
	
	\section{Discussion} \label{sec:disc}
	
	While measurements of the profile shape of Ly$\alpha$ have been demonstrated to correlate with \fesc{} at low redshift \citep[e.g.,][]{izotovLowredshiftLymanContinuum2018}, these results must be validated at high redshift before extrapolating to the reionization era, and must be considered in context of what is feasibly observable at $z\gtrsim5.5$ when the IGM is significantly more neutral. Here, we discuss our results in context of those from low-redshift ($z\sim0.3$) surveys, and attempt to connect our results to observations of the earliest galaxies, including recent results from \textit{JWST}.
	
	\subsection{Comparison to low redshift surveys} \label{sec:compare_lowz}
	
	The connections between \fesc{} and galaxy properties at $z\sim3$ provide key insights into the physics of ionizing-photon escape, and help ground assumptions of \fesc{} for reionization-era galaxies. We found that galaxies with red-side peaks more offset from systemic (thus, higher \vlyar{}) did not have significantly-different \fesc{} values than those with lower \vlyar{}. We do find a significant difference in the IGM-corrected ionizing to non-ionizing flux ratio \fout{} across bins of \vlyar{}, which is less model dependent measurement of the escape of ionizing radiation. Conversely, we find galaxies with larger velocity differences between the peak of Ly$\alpha$ emission and low-ionization absorption absorption (thus, higher \vlma{}) do not have significantly-different \fesc{} or \fout{} values than those with lower \vlma{}. 
	
\begin{figure*} %
		\centering
		\includegraphics[width=\textwidth]{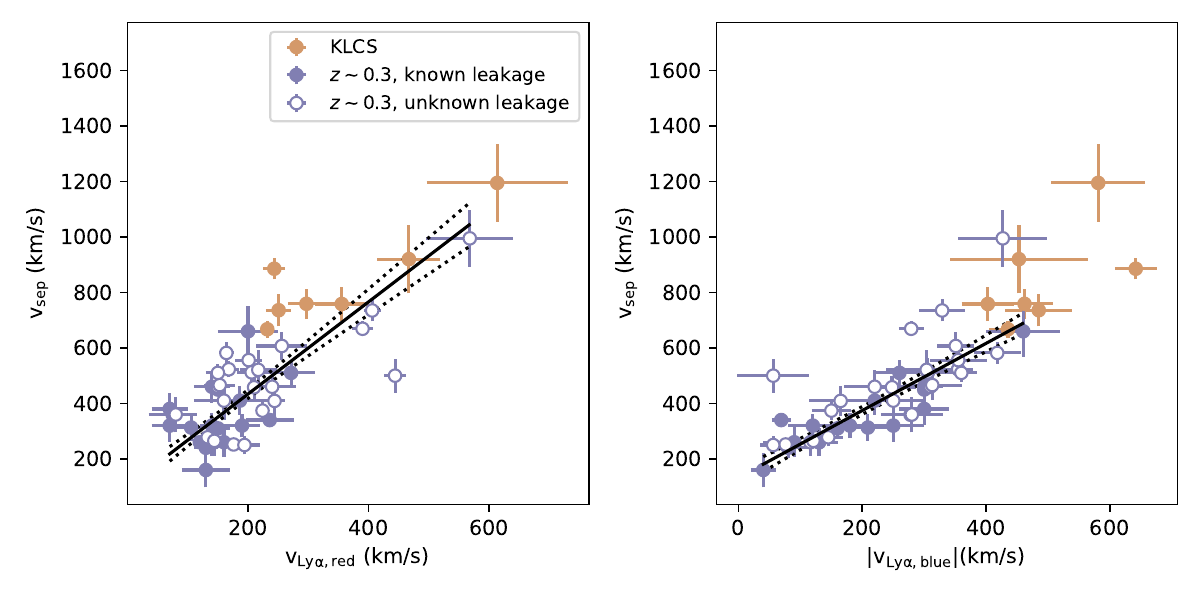}
		\caption{Comparison between the the velocity separation between the peaks of \lya{} emission (\vsep{}) and the velocity offset of different peaks of \lya{} emission and systemic, both at low redshift and this work. Measurements of galaxies at $z\sim0.3$, primarily Green Peas, are sourced from \citet{yangLyaProfileDust2017,gazagnesOriginEscapeLyman2020}; and \citet{fluryLowredshiftLymanContinuum2022}. Measurements presented in \citet{gazagnesOriginEscapeLyman2020} include objects originally published in \citet{pettiniOIIINIIAbundance2004,izotovGreenPeaGalaxies2011,izotovDetectionHighLyman2016,izotovEightCentLeakage2016,izotovLowredshiftLymanContinuum2018,izotovJ11542443Lowredshift2018,leithererDirectDetectionLyman2016}; and \citet{chisholmGalaxiesThatLeak2017}.
			\textbf{Left panel: } Comparison between \vsep{} and the velocity offset of the red side of \lya{} emission and systemic (\vlyar{}). Both KLCS and $z\sim0.3$ samples show a positive correlation between \vsep{} and \vlyar{}, while KLCS objects show higher \vsep{} and \vlyar{} measurements on average than the $z\sim0.3$ objects. A linear relationship fit to the $z\sim0.3$ data is shown as a black, dotted line. The 95\% confidence region for this fit is shown with dashed, black lines.
			\textbf{Right panel: } Comparison between \vsep{} and the velocity offset of the blue side of \lya{} emission and systemic (\vlyab{}). A linear relationship fit to the $z\sim0.3$ data is shown as a black, solid line. The 95\% confidence region for this fit is shown with dashed, black lines. The $z\sim0.3$ \vlyab{}-\vsep{} relation is tighter than that of \vlyar{}-\vsep{}.}
		\label{fig:vlya_lowz}
	\end{figure*}

	In the $z\sim0.3$ Universe, LyC radiation from star-forming galaxies is observable using the \textit{HST}-COS spectrograph and is fully unobscured thanks fewer absorbers in the IGM \citep[e.g.,][]{weymannHubbleSpaceTelescope1998}. Searches for LyC leakers have revealed an anti-correlation between \fesc{} and \lya{} peak separation, \vsep{}, often presented as the strongest tracer of \fesc{} within local reionization-era analogs \citep{izotovDetectionHighLyman2016,izotovLowredshiftLymanContinuum2018,izotovLymanContinuumLeakage2021,fluryLowredshiftLymanContinuum2022,fluryLowredshiftLymanContinuum2022a}. Considering the apparent correlation observed between \vsep{} and \vlyar{} in the KLCS, it is reasonable that we found a significant (1$\sigma$) correlation between \fout{} and \vlyar{} at $z\sim3$. We contrast the \vsep{}-\vlyar{} relationship in the KLCS and that from $z\sim0.3$ Green Peas in literature in the left panel of Figure \ref{fig:vlya_lowz}. A correlation between \vsep{} and \vlyar{} is found in both redshift samples, although KLCS galaxies have larger \vlyar{} and \vsep{} values. We note that \lya{} photons adopt the velocities of neutral Hydrogen in the CGM in order to escape their source galaxy \citep{steidelStructureKinematicsCircumgalactic2010}, and the amplitudes of typical outflow velocities in the CGM are mass dependent, such that more massive galaxies tend to have faster outflows \citep{weinerUbiquitousOutflowsDEEP22009,martinDemographicsPhysicalProperties2012,chisholmSCALINGRELATIONSWARM2015}. Given that the Green Peas from the literature are roughly an order of magnitude lower in \mstar{} than our KLCS LBGs, it is expected that our velocity measurements are higher overall. 
	We perform a orthogonal distance regression on the $z\sim0.3$ data using the python package \textsc{scipy} and find the relation
	\begin{equation} \label{eqn:vsep_vlyar}
		\vsep = 1.133 \times \vlyar + 164.5\:\textrm{km}\:\textrm{s}^{-1}.
	\end{equation}
	We display this best-fit relation and the corresponding 95\% confidence region as a solid and dashed black lines, respectively, in the left panel of Figure \ref{fig:vlya_lowz}. By comparing to the average trend at low redshift, we see a offset to larger \vsep{} values at fixed \vlyar{} at $z\sim3$ when compared to the lower-redshift data. Stated in a different way, this difference indicates that our sample has \lya{} profiles with larger blue-side velocity offsets (i.e., more negative) at fixed red-side velocity offset when compared to low-redshift Green Peas. This difference appears to indicate that the shape of \lya{} profiles may not be structured in the same way across redshift. Unfortunately, this comparison can only be made for a highly limited subsample within the KLCS: this figure contains the seven galaxies with double-peaked \lya{} emission profiles sufficiently resolved in the $R\sim800-1400$ spectra and have follow up observations of rest-optical nebular lines. Given the difference in spectral resolution between the two samples, we discuss the effects of spectral resolution on our velocity measurements in Appendix \ref{sec:app}. A larger sample of resolved blue \lya{} peaks for KLCS objects would allow for a more robust test of this evolution.
	
\begin{figure*} %
		\centering
		\includegraphics[width=\textwidth]{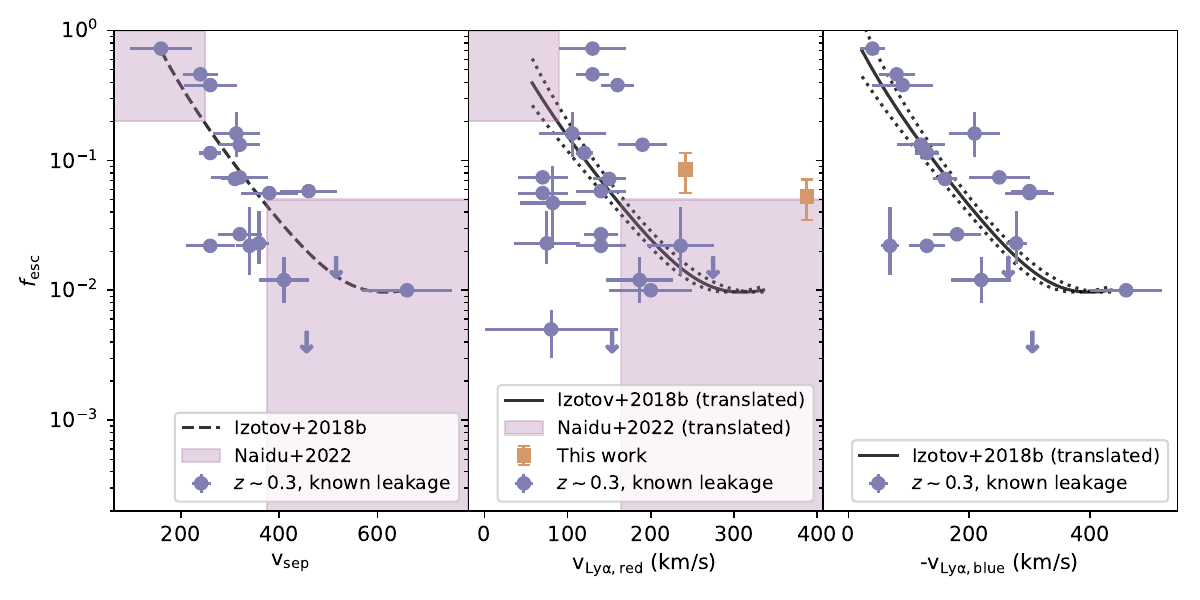}
		\caption{
			\textbf{Left panel:} Comparison between \fesc{} and \vsep{} within $z\sim0.3$ LyC leakers from the literature. These objects are the same as those with known leakage shown in Figure \ref{fig:vlya_lowz}. The best-fit inverse power law from \citet{izotovLowredshiftLymanContinuum2018} is displayed as a dashed, black line. We did not bin the KLCS as a function of \vsep{} as there were too few objects with resolved, double-peaked \lya{} structure. The inferred range of \fesc{} based on \vsep{} from  \citet{naiduSynchronyProductionEscape2022} is displayed as shaded, purple region.
			\textbf{Middle panel:} Measured \fesc{} values for bins of \vlyar{} within the KLCS, alongside $z\sim0.3$ leakers from literature. We translate the \fesc{}-\vsep{} trend of \citet{izotovLowredshiftLymanContinuum2018} into a \fesc{}-\vlyar{} trend using Equation \ref{eqn:vsep_vlyar}, which is derived from $z\sim0.3$ objects. The 95\% confidence interval computed for Equation \ref{eqn:vsep_vlyar} is also translated and is displayed via a dashed line. We also translate the \fesc{} criteria of \citet{naiduSynchronyProductionEscape2022} into \vlyar{} space using the same equation.
			We see a larger \fesc{} at fixed \vlyar{} within the KLCS in comparison to literature $z\sim0.3$ LyC leakers. We also find a significant discrepancy between the \fesc{} of the \vlyarl{} KLCS subsample and the translated \fesc{} criteria of \citet{naiduSynchronyProductionEscape2022}.
			\textbf{Right panel:} \fesc{} vs. \vlyab{} for $z\sim0.3$ LyC leakers from the literature. The solid line is a translated version of the \fesc{}-\vsep{} relation in the left panel, using the linear relation displayed in the right panel of Figure \ref{fig:vlya_lowz} and its corresponding confidence interval. \vlyab{} is a better predictor of \fesc{} than \vlyar{} within the $z\sim0.3$ sample.
		}
		\label{fig:fesc_lowz}
	\end{figure*}

	We expand upon potential differences between low- and high-redshift \lya{} profiles by displaying low-redshift data for both \fesc{} vs. \vsep{} and \fesc{} vs. \vlyar{} in Figure \ref{fig:fesc_lowz}, alongside our results found in bins of \vlyar{}. Considering our small sample size of objects with \vsep{} measurements in the KLCS, we were unable to make bins of sufficient size to precisely constrain \fesc{} as a function of \vsep{} at $z\sim3$. In the left panel of Figure \ref{fig:fesc_lowz}, we show the \fesc{}-\vsep{} relation reported in \citet{izotovDetectionHighLyman2016} alongside published $z\sim0.3$ leakers from the literature, including $z\sim0.3$ Green Peas without LyC observations from \citet{yangLyaProfileDust2017}. We translate this trendline into \vlyar{} space by using Equation \ref{eqn:vsep_vlyar}; the translated relation is displayed as a black, solid line in the middle panel of Figure \ref{fig:fesc_lowz}. At $z\sim0.3$, \fesc{} is much more tightly correlated with \vsep{} than with \vlyar{}. This result appears to indicate that \vlyar{} is a substandard indicator for \fesc{} as compared to \vsep{}. We also find that \vlyar{} is a substandard indicator of \fesc{} as compared to \wlya{} in Figure \ref{fig:fesc}, although correlation is still shown between \fout{} and \vlyar{}. At \vlyar{} values where direct comparison can be made across redshift, we find higher \fesc{} values in the KLCS at fixed \vlyar{} when compared to $z\sim0.3$ leakers. Critically, it appears that relationships between \fesc{} and \lya{} profile shape may not be invariant across redshift. The \klya{} sample has a median $\vlyar=323$\:km\:s$^{-1}$, which would correspond to $\vsep=576$\:km\:s$^{-1}$ according to Equation \ref{eqn:vsep_vlyar}. Using the \fesc-\vsep{} relation fit in \citet{izotovLowredshiftLymanContinuum2018}, this \vsep{} value would correspond to $\fesc\sim0.01$, significantly different from the full-sample measurement of $\fesc=0.06\pm0.01$. At $z\sim3$, it appears that \fesc{} is higher for a given \vsep{} than at $z\sim0.3$. We also note that the while the $z\sim0.3$ samples are lower in stellar mass, they are also generally lower in UV luminosity and higher in SFR as compared to the KLCS, which would not necessarily suppress \fesc{} in these samples \citep{pahlConnectionEscapeIonizing2023}. Furthermore, as discussed in \citet{fluryLowredshiftLymanContinuum2022a}, KLCS composites have higher \fesc{} at fixed \wlya{} than the $z\sim0.3$ objects within the LzLCS survey.
	We therefore exercise caution in using \fesc{} determinations at high redshift (i.e., $z>2$) built directly from \lya{} properties of low ($z\sim0.3$) redshift analogs, discussed further in Section \ref{sec:naidu}.
	
	When comparing \fesc{} estimates across different works, it is necessary to examine any differences in underlying model assumptions. All \fesc{} estimates displayed in Figure \ref{fig:fesc_lowz} assume an intrinsic ionizing flux based on stellar population fits to a galaxy's spectrum, which is subsequently compared to the observed ionizing spectrum in order to compute \fesc{}. Choice of model SED necessarily changes the estimation of \fesc{}. In this work, we used BPASS SPS models, in contrast to the Starburst99 models assumed by the low-redshift LyC surveys discussed in this section \citep{leithererStarburst99SynthesisModels1999}. BPASS, which includes prescriptions for binary stars, yields larger values of intrinsic $f_{900}$ than Starburst99, thus our KLCS galaxies will have comparatively-lower \fesc{} values from this model choice alone \citep{chisholmConstrainingMetallicitiesAges2019}. 
	Treatment of dust within spectral fits similarly affects \fesc{} estimates. In our \fesc{} modeling discussed in Section \ref{sec:spec}, we assumed an SMC dust attenuation curve, and that the ISM within KLCS galaxies include ``holes" free of both neutral gas and dust. The $z\sim0.3$ \fesc{} estimates discussed in this section instead assume a \citet{reddySPECTROSCOPICMEASUREMENTSFARULTRAVIOLET2016} dust attenuation curve, and that the dust is situated in a uniform screen. If these model choices were applied to our analysis of KLCS objects, our reported \fesc{} values would be elevated \citep{steidelKeckLymanContinuum2018}. Considering that our model choices only shift our \fesc{} estimates to comparatively-lower values, we conclude that the \textit{increased} \fesc{} at fixed Ly$\alpha$ profile shape seen within the KLCS is not driven by inconsistent methodology.
	
	\newpage
	
	\subsection{Red vs. blue Ly$\alpha$ peaks}
	
	Considering the apparent relationship between \vsep{} and \vlyar{} across redshift, one would expect each to modulate the escape fraction of ionizing photons in a similar way. The tenuous trend between \fesc{} and \vlyar{} within the $z\sim0.3$ high-redshift analogs motivates a closer examination of which physical properties of the neutral-phase ISM and CGM are probed by \vsep{} and \vlyar{}. 
	Additionally, in the reionization era, any \lya{} emission bluewards of systemic will be resonantly absorbed by optically thick neutral gas in the nearby IGM \citep[e.g.,][]{masonMeasuringPropertiesReionized2020}, making the observation of double-peaked structure exceedingly difficult. Understanding any physical differences between \vsep{} and \vlyar{}, and analogously, \vlyab{} and \vlyar{}, is of critical importance when interpreting \lya{} profiles during the epoch of reionization in the context of escaping ionizing radiation.
	
	We show \vsep{} vs. \vlyab{} for the KLCS and $z\sim0.3$ samples in the right panel of Figure \ref{fig:vlya_lowz}, and see a tighter correlation as compared to that of \vsep{} and \vlyar{} across both redshift samples. This tighter correlation was reported in previous analyses of $z\sim0.3$ Green Peas \citep{henryLyaEmissionGreen2015,verhammeLymanaSpectralProperties2017,orlitovaPuzzlingLymanalphaLine2018}.  We display the $z\sim0.3$ \fesc{}-\vlyab{} correlation from literature sources in the right panel of Figure \ref{fig:fesc_lowz}, and see a similar correlation as that with \vsep{}, in contrast to the relative lack of correlation with \vlyar{}. Again, due to a lack of a large sample of KLCS galaxy measurements with \vlyab{}, these relationships remain untested at high redshift. However, at lower redshift it is clear that the configuration of the neutral-phase ISM and CGM that concerns ionizing-photon escape are more tied to the observable parameters of \vlyab{} and \vsep{}, rather than \vlyar{}.
	
	Simple radiative transfer models of Ly$\alpha$ feature the neutral-phase CGM distributed in a homogeneous shell around a source. In these models, the escape fraction of ionizing radiation is related to the column density of neutral hydrogen (N$_{\rm HI}$), which is optically thin to LyC radiation. In these models, both \vsep{} and \vlyar{} are modulated by N$_{\rm HI}$, although small \vlyar{} is not a sufficient condition for identifying a LyC leaker. Lower values of \vlyar{} may also be driven by high outflow velocities, rather than higher \lya{} and LyC escape fractions \citep{verhammeUsingLymanDetect2015}. Our measurements of \vlma{} contrast the position of the red-side peak of \lya{} to that of low-ionization metal absorption lines, which are produced in the intervening material in the ISM/CGM and directly encode their kinematics. We find a lack of correlation between \fesc{} and \vlma{} within the KLCS, which supports the idea that greater outflow velocities do not significantly affect measured \fesc{} values. More systemic redshift measurements are needed to directly test the correlations between \fesc{} and LIS velocities within the KLCS.
	Nonetheless, LIS lines likely do encode information about the covering fraction of neutral material and thus \fesc{}: \citet{saldana-lopezLowRedshiftLymanContinuum2022} have presented empirical calibrations that use the \textit{depth} of LIS lines and dust attenuation measurements to predict the escape fractions of LzLCS galaxies \citep[see also][]{reddyConnectionReddeningGas2016}.
	
	It's likely that the multiphase ISM in $z\sim3$ galaxies is more complex and dynamically perturbed than homogeneous shells. Our modeling of composite spectra assumes that clear channels in an optically-thick neutral-phase medium allow \lya{} and LyC photons to escape unabated. This scenario is supported observationally by the connections between \lya{} equivalent width, LIS absorption strength, and E(B-V) found in $z\sim2-5$ galaxy surveys \citep[e.g.][]{shapleyRestFrameUltravioletSpectra2003,jonesKeckSpectroscopyFaint2012,duRedshiftEvolutionRestUV2018,pahlUncontaminatedMeasurementEscaping2021}.
	Indeed, in \citet{kakiichiRadiationHydrodynamicsTurbulent2021a}, radiation hydrodynamics simulations indicate that galaxies with highly-asymmetric Ly$\alpha$ profiles, such as those in the KLCS, feature both optically-thick and thin channels that govern the escape of LyC photons. These types of profiles are ubiquitous at high redshift \citep{steidelStructureKinematicsCircumgalactic2010,kulasKinematicsMultiplepeakedLya2012,hashimotoCloseComparisonObserved2015,trainorSpectroscopicPropertiesLyaEmitters2015}. 
	Modeling pursued in \citet{orlitovaPuzzlingLymanalphaLine2018} notes the puzzling differences between \vlyar{} and \vlyab{} in the context of poor fits when assuming a homogeneous, expanding shell of neutral Hydrogen. They find that by including systemic redshifts in their fits, the spherical-shell fit fails, as objects with weak blue-side \lya{} peaks do not have symmetric \lya{} profiles across systemic.
	Other studies of the radiative transfer of \lya{} explore the differences between \vlyar{} and \vlyab{}.
	In \citet{gronkeLymanaSpectraMultiphase2016}, the authors utilize Monte-Carlo radiative transfer simulations of \lya{} through a clumpy medium and find the physical parameters that affect the emergent profile shape. They find that \vlyar{} is sensitive to clump covering fraction and clump radial velocity, while \vlyab{} is more sensitive to the column density of the static, hot inter-clump medium.
	
	Under an assumption of a heterogeneous medium, the red-side \lya{} profile will not directly reflect properties of the foreground gas. \lya{} photons redward of systemic primarily take paths that backscatter off of outflowing material behind the galaxy, while those blueward of systemic are scattered off resonance by gas \citep[e.g.,][]{steidelStructureKinematicsCircumgalactic2010}. This fact alone may explain why \fesc{} is less sensitive to \vlyar{} than \vsep{} or \vlyab{}, as gas in the background does not attenuate LyC photons traveling in the line-of-sight direction. 
	Considering that viewing angle is expected to strongly affect emergent \lya{} profiles \citep[e.g.,][]{blaizotSimulatingDiversityShapes2023}, it stands to reason that sample-averaged statistics may smooth out these variations, and \vlyar{} is still predictive of \textit{average} ionizing-photon escape, as in the KLCS. 
		
		\subsection{The \fesc{} assumptions of \citet{naiduSynchronyProductionEscape2022}} \label{sec:naidu}
		
		The primary goal of this work was to test the connection between \fesc{} and \lya{} profile shape in the $z\sim3$ Universe, and whether trends found at lower redshift hold. While the KLCS does not currently have the spectral resolution to discern double-peak structure for the majority of the sample, detailed spectral analyses of higher-resolution \lya{} profiles at intermediate redshifts have been performed, like those of the XLS-$z$2 survey \citep{mattheeXSHOOTERLymanSurvey2021}. We can use the \lya{} profiles of the KLCS, which have the added benefit of direct LyC measurements, to add indirect \fesc{} constraints to the objects in this survey.
		
		The XLS-$z$2 survey features 35 $z\sim2$ LAEs, selected primarily via narrow-band imaging and targeted with the X-SHOOTER spectrograph on VLT with spectral resolution of $R\sim4000$. Systemic redshifts were also recovered for 33/35 of the sources. Using criteria empirically motivated from the \lya{} profiles of LyC leakers, \citet{naiduSynchronyProductionEscape2022} presented a framework for predicting \fesc{} using the \lya{} profiles of XLS-$z$2 objects alone. The resulting criteria were based on measurements from LyC leakers at $z\sim0.3$, as well as a few individually-leaking objects found at $z\sim2-4$ (although individual LyC detections at high redshift have highly-uncertain \fesc{}). The criteria were two-fold: objects with low \vsep{} or high flux densities at systemic (\fcen{}) were assumed to have $>20\%$ \fesc{}, while objects with high \vsep{} and low \fcen{} were assumed to have $<5\%$ \fesc{}. The \vsep{} criteria are displayed in shaded regions in the left panel of Figure \ref{fig:fesc_lowz}. By introducing criteria based on \vsep{}, the sample also demonstrated a connection between assumed \fesc{} and \vlyar{}. Composite spectra of the low- and high-\fesc{} XLS-$z$2 galaxies featured different \vlyar{}: those with assumed $\fesc>20\%$ had $\vlyar=106\pm3$\:km\:s$^{-1}$, while those with $\fesc<5\%$ had $\vlyar=254\pm4$\:km\:s$^{-1}$.
		
		We attempt to test the \fesc{} assumptions of \citet{naiduSynchronyProductionEscape2022} directly with the KLCS by translating the \vsep-based \fesc{} criteria into one based on \vlyar{}. Using Equation \ref{eqn:vsep_vlyar} and assuming a direct relationship between the two variables, the ``high escape" objects would have $\vlyar<89$\:km\:s$^{-1}$, while the ``low escape" objects would have $\vlyar>164$\:km\:s$^{-1}$. The translated criteria are displayed as shaded regions in the middle panel of Figure \ref{fig:fesc_lowz}.  The KLCS objects span a dynamic range of \vlyar{} that allow a comparison to the $\fesc<5\%$ region: the \vlyarl{} composite has $\fesc=0.08\pm0.03$, which lies outside of the region.
		The ``low escape" spectral stack of \citet{naiduSynchronyProductionEscape2022} can also be directly compared to this \vlyarl{} KLCS stack. The X-SHOOTER composite spectrum has $\vlyar=254\pm4$\:km\:s$^{-1}$, $\wlya=61\pm3$\AA{}, and an assumed $\fesc<5\%$. In comparison, the \vlyarl{} stack has a median $\vlyar=242\pm14$\:km\:s$^{-1}$ and $\wlya=21\pm7$. The \vlyarl{} stack, despite having lower \wlya{} than that of the ``low escape" stack of \citet{naiduSynchronyProductionEscape2022}, has a significantly larger \fesc{} value than that assumed when using the \lya{}-based criteria in \citet{naiduSynchronyProductionEscape2022}. It is likely that \fesc{} measured for the \vlyarl{} stack would be even higher if the \wlya{} were comparable to the ``low escape" stack of \citet{naiduSynchronyProductionEscape2022},  considering the direct connection between \fesc{} and \wlya{} ubiquitous in $z\sim2-4$ LyC samples \citep{steidelKeckLymanContinuum2018,marchiLyaLymanContinuumConnection2018,begleyVANDELSSurveyMeasurement2022}. While we cannot directly test the \fesc{} assumptions of \citet{naiduSynchronyProductionEscape2022} without higher-resolution spectra of the KLCS, these discrepancies indicate that \fesc{} assumptions based on \lya{} profile shape may not be consistent across all samples, especially across redshift and/or galaxy properties such as stellar mass, UV luminosity and SFR. From the comparison presented in this section, it appears that \fesc{} values inferred from low-$z$ scaling relations will be underestimated.
		
		\subsection{Ly$\alpha$ profiles during reionization}
		
		The connections between \fesc{} and galaxy properties at $z\sim3$ can indicate the most appropriate assumptions for \fesc{} at even higher redshifts, during the epoch of reionization. We find that galaxies with red-side peaks more offset from systemic (thus, higher \vlyar{}) tend to have  lower \fout{} values, although they did not have significantly-different \fesc{} values. We also find that, at fixed \vlyar{} (and implied \vsep{}), \fesc{} is higher at $z\sim3$ than at $z\sim0.3$.
		Nonetheless, \vlyar{} may be the best proxy of \vsep{} actually observable for galaxies in the reionization era. In simulations of \lya{} transmission fractions at $z\sim6$, typical sightlines through the neutral-phase IGM feature feature complete attenuation of Ly$\alpha$ flux below 1216\AA{} \citep{laursenIntergalacticTransmissionIts2011}.
		
		Despite the difficulty of observing blue-side \lya{} flux in the epoch of reionization, a handful of $z\gtrsim6$ galaxies have been observed with this signature \citep{huULTRALUMINOUSLYaEMITTER2016,songailaComplexLyaProfiles2018,bosmanThreeLyaEmitting2020}. The mere observability of blue-side \lya{} indicates that these photons are traveling through highly-transparent and unique sightlines through the IGM, and these objects are residing in ionized bubbles \citep{masonMeasuringPropertiesReionized2020}. While such signatures are useful for examining these rare objects and probing the topology of reionization, the vast majority of galaxies will not have observed blue-side \lya{} flux.
		
		In contrast, the red-side peak of \lya{} and associated nebular emission lines required to constrain \zsys{}, and thus \vlyar{}, are observable at extreme redshifts. A limited number of \vlyar{} measurements have been made for $z>6$ galaxies using C~\textsc{iii}] as a probe of \zsys{} \citep{starkSpectroscopicDetectionIV2015,starkLyaIIIEmission2017,hutchisonNearinfraredSpectroscopyGalaxies2019}. More recently, surveys targeting [C~\textsc{ii}] and [O~\textsc{iii}] with ALMA have found further success \citep[e.g.,][]{cassataALPINEALMACIISurvey2020,endsleyREBELSALMASurvey2022}.
		With the advent of JWST/NIRSpec, rest-optical nebular emission lines are more readily recoverable at $z>6$, and samples with \vlyar{} measurements are growing \citep{tangJWSTNIRSpecSpectroscopy2023}. These objects have \vlyar{} between $\sim100$\:km\:s$^{-1}$ and $\sim800$\:km\:s$^{-1}$, with $\wlya<100$\AA{} and $-23<$M$_{\rm UV}<-19$. Our \fesc{} estimates as a function of both \vlyar{} and \wlya{} represent the first comparison sample to these objects drawn from a $z\sim3$ sample. Based on \lya{} alone, objects such as REBELS-23 \citep[$\vlyar=227$\:km\:s$^{-1}$, $\wlya=14$\AA{};][]{endsleyREBELSALMASurvey2022} may have $\fesc\sim0.08$, considering its similarity to our \vlyarl{} subsample.
		
		Direct comparison to \lya{} profiles at $z>6$ has some caveats. At the neutral fractions of the IGM during reionization, even the blue side of the red \lya{} peak begins to be attenuated, which is more prominent closer to systemic \citep{laursenIntergalacticTransmissionIts2011}. This attenuation will bias \vlyar{} to higher values as the \lya{} profiles become asymmetrically modulated. Additionally, \lya{} profiles that have strong emission significantly redward of systemic will be preferentially observed, preventing an unbiased view into the average velocity offsets of different populations of reionization-era galaxies. Despite these caveats, there is promising evidence that the \lya{} profiles of UV bright populations may not be fully attenuated by the neutral IGM at $z>6$. The \wlya{} distribution in M$_{\rm UV}<-21$ galaxies appears not to decline from $6<z<8$ \citep{endsleyMMTSpectroscopyLymanalpha2021,jungCLEARBoostedLya2022,roberts-borsaniNatureNurtureComparing2023}, and \lya{} appears to be ubiquitous in this population at $z\sim7.5-9$ \citep[e.g.,][]{starkLyaIIIEmission2017,larsonSearchingIslandsReionization2022}. Thus, for luminous galaxies in particular, the strength and shape of \lya{} may be a valid indicator of \fesc{} at $z\gtrsim6$.

		\section{Summary} \label{sec:summary}
		
		In this work, we investigate the connections between \lya{} profile shape and ionizing-photon escape at high redshift, as a complement and comparison to existing analyses at $z\sim0.3$ \citep[e.g.,][]{izotovDetectionHighLyman2016,izotovLowredshiftLymanContinuum2018,izotovLymanContinuumLeakage2021,fluryLowredshiftLymanContinuum2022,fluryLowredshiftLymanContinuum2022a}.
		To this end, we utilized  KLCS galaxies, for which $R\sim800-1400$, rest-UV spectroscopy is available. These spectra cover both the \lya{} spectral feature and the LyC spectral region, with sufficient depth to sensitively constrain \fesc{} in composite spectra. By making quantitative measurements of \lya{} and binning the KLCS as a function of these properties, we explored the connections between the velocity structure of \lya{} and the escape of ionizing radiation in $z\sim3$ star-forming galaxies. Our main conclusions are as follows:
		
		\begin{enumerate}
			\item We find positive correlations between the velocity separation between the red-side peak of \lya{} and systemic redshift \vlyar{} and the \lya{} peak separation \vsep{} within the KLCS, but we lack the sample size to quantify the significance of this trend. We find a significant correlation between \vlyar{} and and velocity separation between the red-side peak of \lya{} and the low-ionization interstellar redshift \vlma{}. When comparing our measurements to samples of $z\sim0.3$ Green Peas from the literature, we find an evolution in the \vsep{}-\vlyar{} relationship with redshift. Galaxies at $z\sim3$ from the KLCS have larger \vsep{} at fixed \vlyar{}, indicating that \lya{} profile shapes potentially evolve with redshift and/or stellar mass.
			\item We find a significant (1$\sigma$), negative correlation between the ionizing to non-ionizing flux ratio \fout{} and \vlyar{} within the KLCS, found across two bins of \vlyar{}. We find no significant differences for any measure of ionizing-photon escape across two bins of \vlma{}. The connection between \fout{} and \vlyar{} is supported by radiative transfer models that connect increasing covering fractions of neutral gas to larger velocity shifts that \lya{} photons must adopt to escape from a galaxy \citep[e.g.,][]{steidelStructureKinematicsCircumgalactic2010}.
			We contrast our \fesc{} measurements in two bins of \vlyar{} to that of literature LyC leakers at $z\sim0.3$, and find larger \fesc{} at fixed \vlyar{} within our higher-redshift sample. From this evolution, we infer that while not only are \lya{} profiles different across redshift and stellar mass, \lya{} profile shape measurements such as \vlyar{} appear to trace escaping ionizing radiation in a different manner. Building on this point, we perform a linear fit between \vsep{} and \vlyar{}, and we find larger \fesc{} at fixed \vsep{} than those predicted by $z\sim0.3$ data. We further conclude that these differences could not be caused by inconsistent model assumptions.
			\item We compare the correlations between \fesc{} and \vlyar{}, \vsep{}, and \vlyab{} by compiling results for $z\sim0.3$ LyC leakers in literature. We find that \fesc{} correlates much more strongly with \vsep{} and \vlyab{} than with \vlyar{} at $z\sim0.3$, despite our measured correlation between \fout{} and \vlyar{} within the KLCS. We infer that the heterogeneity of neutral-phase gas around LyC leakers, which manifests in a large viewing-angle dependence on both \lya{} profile shape and \fesc{} \citep[e.g.,][]{verhammeLymanaEmissionProperties2012,behrensInclinationDependenceLymana2014,smithPhysicsLymanEscape2019,smithPhysicsLymanaEscape2022,smithTHESANProjectLymana2022,blaizotSimulatingDiversityShapes2023}, will scatter the relationship between \fesc{} and \vlyar{} in individual objects, considering the red side of \lya{} is primarily concerned with gas in the background of a galaxy. Within the KLCS, we effectively averaged across many viewing angles by creating composite spectra of subsamples binned as a function of \vlyar{}, thus probing the average neutral-gas configuration of galaxies within each stack.
			\item We compare our low-\vlyar{} bin to that of the ``low escape" stack of the XLS-$z$2 survey presented in \citet{naiduSynchronyProductionEscape2022}, in order to test \fesc{} assumptions primarily calibrated using local samples. Our KLCS composite constructed from objects with the lowest \vlyar{} measurements had an estimated $\fesc=0.08\pm0.03$, greater than the assumed $\fesc<0.05$ in a XLS-$z$2 composite with comparable \vlyar{}. While we lack the higher resolution spectra to directly test their \vsep{} and central-flux-based criteria, the discrepancy between the assumed \fesc{} at fixed \vlyar{} indicates that further investigation  is needed into the \lya{} profile shapes of LyC leakers across redshift, at larger dynamic ranges of \vlyar{} and \vsep{} than is probed by the KLCS.
		\end{enumerate}
		
		In the reionization epoch, any \lya{} flux blueward of systemic emitted from typical galaxies is entirely attenuated due to the neutral-phase IGM. These results, presented as a function of red-side \lya{} profile shape, are critical for developing new \lya{}-based criteria usable for $z>6$ galaxies. In addition, we've demonstrated that aspects of the \lya{} profile and it's effects on ionizing-photon escape may evolve with redshift and stellar mass. These analyses, performed at the some of the highest redshift possible, provide important new insights into the mechanics of ionizing-photon escape during reionization. Larger samples with high-resolution, rest-UV spectra covering \lya{}, coupled with direct LyC measurements, across larger dynamic ranges of galaxy properties are necessary to build on these results and construct a complete suite of diagnostics to predict \fesc{} of reionization-era galaxies.
		\linebreak
		\linebreak
		The authors acknowledge helpful conversations with John Chisholm and Drew Newman during the drafting of this manuscript.
		AJP was generously supported by a Carnegie Fellowship through the Carnegie Observatories while conducting part of this work.
		We acknowledge support from NSF AAG grants 0606912, 0908805, 1313472, 2009313, 2009085, and 2009278.
		Support for program HST-GO-15287.001 was provided by NASA through a grant from the Space Telescope Science Institute, which is operated by the Associations of Universities for Research in Astronomy, Incorporated, under NASA contract NAS5-26555. 
		CS was supported in part by the Caltech/JPL President's and Director's program.
		We wish to extend special thanks to those of Hawaiian ancestry on
		whose sacred mountain we are privileged to be guests. Without their generous hospitality, most
		of the observations presented herein would not have been possible.
		\appendix
		\section{Feasibility of recovering trends between \fesc{} and Ly$\alpha$ profile shape for different sample sizes} \label{sec:app2}
		In \citet{pahlSearchingConnectionIonizingphoton2022}, we performed simulations to determine the minimum sample sizes required to confidently recover trends between \fesc{} and galaxy properties within the KLCS. 
		We used the trend between \fout{} and \wlya{} as a measure of the reliability of smaller samples, considering the ubiquity of trends between ionizing-photon escape and \wlya{} found within $z\sim3$ LyC surveys \citep{marchiNewConstraintsAverage2017,marchiLyaLymanContinuumConnection2018,fletcherLYMANCONTINUUMESCAPE2019,steidelKeckLymanContinuum2018}.
		\fout{} is defined as the ratio between $f_{900}$ and the average flux density in the non-ionizing UV continuum ($1475-1525$\AA{}, $f_{1500}$) after performing an IGM+CGM transmission correction.
		It is a more model-independent measurement of escaping ionizing light than \fesc{}, since it is not based on any stellar-population-dependent assumption of the intrinsic production rate of ionizing radiation.
		We performed simulations by drawing random samples from the full KLCS of a fixed sample size, binning the objects into two \wlya{} subsamples, and determining whether the resulting composite spectra feature a 1$\sigma$ difference between their estimated \fout{} values. 
		The \citet{pahlSearchingConnectionIonizingphoton2022} simulations were performed in two different modes: one assuming preferential selection of LyC detections, and one with entirely random selection. The preferential selection mode included all 13 KLCS galaxies individually detected in LyC for each random draw. This mode was relevant for subset of galaxies in \citet{pahlSearchingConnectionIonizingphoton2022}, which included all LyC detections. For this work, we consider the preferential LyC detection mode comparable to the \klya{} sample, considering that 8/45 objects are LyC detections, compared to 13/120 for the full KLCS. Both \klma{} samples and \ksep{} samples are more comparable to the random selection mode (6/67 and 1/15 are LyC detections, respectively).
		
		We used the results from these simulations to determine if the sizes of each samples are adequate for investigation of \fesc{} and the respective binned property. In the LyC preferential-selection mode, a sample size of 45 resulted in a recovery of the \fout{}-\wlya{} trend in 92\% of realizations, indicating that \klya{} is appropriate for determining trends between \fesc{} and \vlyar{}. In fully random mode, a sample size of 63 finds the same recovery in 80\% of realizations, while a sample size of 15 recovers the trend in 30\%. We therefore disinclude \ksep{} from analyses of ionizing photon escape.
		
		\section{Effects of spectral resolution on velocity measurements} \label{sec:app}
		Our conclusions herein are based on relatively low resolution ($R\simeq800$ for LRIS-B, $z<3.11$; $R\simeq1400$ for LRIS-R, $z>3.11$) spectra, which results in a small subset of the KLCS with resolved blue-side \lya{} peaks, and thus measurable peak separations. 
		We therefore relied on the velocity offset of the red-side peak of \lya{} within this analysis, and in our comparison to low-redshift data (HST-COS; $R\simeq16000$ at \lya{}) and the survey of LAEs of XLS-$z$2 discussed in \citet{naiduSynchronyProductionEscape2022} (X-SHOOTER, $R\simeq4000$ at \lya{}). Considering the difference in spectral resolution between the KLCS observations originally presented in \citet{steidelKeckLymanContinuum2018} and these surveys, we discuss the effects of spectral resolution on our velocity measurements.
		
		In \citet{verhammeUsingLymanDetect2015}, the authors perform simulations on theoretical \lya{} profiles to understand the effects of broadening from lower-resolution spectrographs on the shape of \lya{}. In the lowest resolution ($R\sim1000$) simulation, the blue peak of \lya{} becomes a bump on the wing of the stronger, red-side component of the \lya{} profile, or simply presents an elongated blue-side tail. Similar profile shapes are observed within the KLCS. Across the different-resolution simulations, the location of the red-side peak of \lya{} is well conserved. The peak shifts only slightly to lower values (by $<50$\:km\:s$^{-1}$), indicating that any bias in \vlyar{} within our sample is minimal.
		
		We can directly quantify the effects of spectral resolution on the position of the red-side peak of \lya{} for a subset of the KLCS. In order to resolve additional double-peaked structure of KLCS galaxies with \lya{} in emission, we have obtained a limited amount of follow up Keck/LRIS spectroscopy with the 1200-line grating on the red side of the dichroic, featuring a spectral resolution of $R\sim2500$. While the sample of KLCS objects with \vsep{} remains modest, we measured the \lya{} red-side peak offsets from systemic, thus \vlyar{}, for 11 KLCS objects with this higher-resolution data following the procedures presented in Section \ref{sec:z_methods}. We calculate the velocity difference between \vlyar{} measured from the low-resolution LRIS data and that measured from the high-resolution data, and find a median $\Delta$v$=+56$\:km\:s$^{-1}$ with a standard deviation of 63\:km\:s$^{-1}$, where the high-resolution data typically features smaller \vlyar{} measurements. This difference represents a small shift compared to the median \vlyar{} of our sample of $319$\:km\:s$^{-1}$, and is comparable to the typical errors on the \vlyar{} measurements of $39\: $km\:s$^{-1}$. Additionally, our results are presented using bins of \vlyar{} all measured from spectra of similar resolution, thus, our result of ionizing-photon escape correlations with \vlyar{} will not be affected by any potential biases from low spectral resolution. Nonetheless, it is worth noting potential discrepancies when comparing the \vlyar{} values of this work and others with higher spectral resolution.


	\end{document}